\documentclass[11pt]{article}
\usepackage{jheppub}
\usepackage{graphicx}
\usepackage{amsmath,amssymb,amstext,amsfonts} 
\usepackage{mathrsfs}
\usepackage[utf8]{inputenc}
\usepackage[a4paper,top=1.89in, bottom=0in, left=2in, right=0in]{geometry}
\usepackage{xcolor}
\usepackage{braket}
\usepackage{array}
\usepackage[bottom]{footmisc}
\usepackage{tikz}
\usepackage{tabularx}
\usepackage{setspace}
\usepackage{enumitem}
\usepackage[numbers]{natbib}
\usepackage{caption}
\usepackage{subcaption}
\usepackage[normalem]{ulem}

\newcommand{\be}{\begin{equation}}
	\newcommand{\ee}{\end{equation}}
\newcommand{\ba}{\begin{eqnarray}}
	\newcommand{\ea}{\end{eqnarray}}

\numberwithin{equation}{section}

\begin{document}
	\title{Phase Structure and Thermodynamic Geometry of STU Black Holes}
	
	\author[1]{Fil Simovic}
	\author[2]{Kenan Scavennec}
	\author[3]{David Kubiznak}
	\author[4]{Robert B. Mann}
	
	\affiliation[1]{Department of Physics and Astronomy, University of Waterloo, 
		Waterloo, ON, N2L 3G1, Canada\\
		School of Mathematical and Physical Sciences, Macquarie University, Sydney, Australia}
	
	\affiliation[2]{D\'epartement de Physique de l'\'Ecole Normale Sup\'erieure, ENS,
		Universit\'e PSL,
		F-75005 Paris, France
	}
	
	\affiliation[3]{Institute of Theoretical Physics, Faculty of Mathematics and Physics,
		Charles University, Prague, V Hole{\v s}ovi{\v c}k{\' a}ch 2, 180 00 Prague 8, Czech Republic}
	
	\affiliation[4]{Department of Physics and Astronomy, University of Waterloo, 
		Waterloo, ON, N2L 3G1, Canada\\
		Perimeter Institute  for Theoretical Physics, 31 Caroline Street North, Waterloo, ON, N2L 2Y5, Canada}
	
	\emailAdd{fsimovic@uwaterloo.ca}
	\emailAdd{kenan.scavennec@ens.psl.eu}
	\emailAdd{david.kubiznak@matfyz.cuni.cz}
	\emailAdd{rbmann@uwaterloo.ca}

	\abstract{We study the thermodynamics of four-dimensional STU black holes, a family of charged solutions of gauged $N=8$, $SO(8)$ supergravity, from the perspective of geometric thermodynamics. We focus on non-rotating solutions carrying pairwise-equal charges, obtaining closed-form expressions for the thermodynamic quantities directly in terms of the physical charges, and use them to determine the equation of state and associated phase structure. We also construct the Ruppeiner geometry for the STU black hole, taking the temperature and thermodynamic volume as fluctuation variables, and study its renormalized curvature as a probe of the black hole microstructure. We find a rich phase structure with two critical points present whenever the charge ratio is sufficiently large, and identify regions in parameter space associated with attractive and repulsive microscopic interactions.}
	

	\maketitle
	
	\section{Introduction}
	
	Thermodynamics has historically provided a means of inferring microscopic structures that could not otherwise be seen, the classical example being the use of the gas laws to formulate statements regarding atomic interactions prior to observation of atoms themselves. Black hole thermodynamics extends this idea to gravity. Although a black hole can be described in terms of general relativity using only a few macroscopic variables, it follows laws that are formally equivalent to those of thermodynamics, with the area of the black hole event horizon acting as its entropy and the surface gravity of the black hole acting as its temperature \cite{bardeen1973,bekenstein1973}. Hawking's demonstration that the event horizon radiates thermally with temperature proportional to its surface gravity turns this analogy into a physical identity \cite{hawking1975}, which then carries an important implication. The usual statistical reasoning associates temperature and entropy with a set of microscopic degrees of freedom which black holes must then possess. Because the laws of black hole thermodynamics follow from general relativity and thermodynamics alone, the laws provide an important consistency test for candidate quantum theories of gravity, as well as a window onto the microscopic structure those theories must ultimately describe \cite{mann2026}.
	
	One of the most direct ways to study these microscopic degrees of freedom is to geometrize thermodynamics itself. This strategy, originating in the differential-geometric reformulations of equilibrium thermodynamics by Gibbs and Carath\'eodory and developed by Hermann, Mrugala, Weinhold, and Ruppeiner, involves representing equilibrium states as points on a manifold whose phase space carries the structure of a contact manifold \cite{weinhold1976,ruppeiner1995}. It is closely tied to information geometry, where the corresponding manifolds are spaces of probability distributions equipped with the Fisher information metric \cite{brody2003}. Of special interest is the  Ruppeiner geometry, constructed from the Hessian of the entropy with respect to the system's extensive variables, whose line element measures the likelihood of fluctuations between nearby equilibrium states. The sign of its scalar curvature is widely believed to reflect the character of the underlying microscopic interactions, being positive for predominantly repulsive interactions and negative for attractive ones, while its magnitude scales with the correlation length and diverges at critical points \cite{ruppeiner1995,ruppeiner2023}. Applied to black holes, the Ruppeiner curvature has served as a probe of the largely inaccessible microstructure of the gravitational field. For the Schwarzschild and Reissner--Nordstr\"om solutions, a suitable renormalization of the curvature scalar recovers features of the effective potential responsible for the coarse-grained thermodynamics \cite{ruppeiner2023}, while for the Reissner--Nordstr\"om-AdS black hole, it seems to provide the first known example in which the close analogy between black hole thermodynamics and Van der Waals fluids \cite{chamblin1999a, chamblin1999b, cvetic1999a, kubiznak2012} breaks down, exhibiting a region of phase space in which the microscopic interactions appear repulsive, with no analog in the Van der Waals system \cite{wei2019a, wei2019b}.

	Such studies are naturally carried out in asymptotically anti-de Sitter spaces. Treating the cosmological constant as a thermodynamic pressure and the ADM mass as an enthalpy promotes the first law to an {extended} form carrying a pressure-volume term \cite{kastor2009, dolan2011a,cvetic2011, dolan2011b}, and the resulting phase space is in many regards fluid-like, supporting Van der Waals-type small-large black hole transitions \cite{kubiznak2012}, and in richer settings, reentrant transitions,  multi-critical points, isolated critical points, or superfluid-like phase transitions  \cite{altamirano2013, altamirano2014a,frassino2014, dolan2014, hu2024,  hennigar2017, tavakoli2022,wu2023a,wu2023b,wu2023c,lu2024,kubiznak2017, mann2025}. Each of these has a dual interpretation through the AdS/CFT correspondence, describing the thermalization of strongly coupled field theories \cite{maldacena1998, witten1998, aharony2000};
	see \cite{karch2015, visser2022, cong2021,  ahmed2023} for recent proposals for the holographic interpretation of extended phase space thermodynamics and their critique \cite{mancilla2025, borsboom2026}. It is in this {\em extended phase space} that we work throughout.

	We apply these tools to an important family of black hole solutions arising in gauged supergravity, the STU black holes \cite{duff1999, cvetic1999b, chong2005} (see also the most recent generalization \cite{ma2026, wu2026}).
	These are generalizations of the charged, rotating Kerr--Newman metric of general relativity, carrying (in addition to mass and angular momentum) up to four independent charges sourced by $U(1)^4$ gauge fields of the theory, which descends from the massless sector of eleven-dimensional supergravity on $AdS_4\!\times\!\mathcal{S}^7$.\footnote{More general solutions of the model, also carrying magnetic and NUT charges, have been constructed in \cite{chow2014}.} The Kerr--Newman and Reissner--Nordstr\"om solutions are recovered in appropriate limits, and the additional charges endow the geometry with a richer causal and thermodynamic structure while leaving its null geodesics integrable, e.g. \cite{cvetic2017}. From a phenomenological standpoint the STU black holes are interesting as they carry enough free parameters to allow for systematic tests of departures from general relativity, yet not so many as to render the space of metrics effectively unconstrained. For the present study we restrict attention to the non-rotating solutions, and for analytic tractability, to the case of two distinct charges, obtained by setting the four charges pairwise equal; see \cite{cvetic2011, caceres2015, sekhmani2023} for previous studies of extended phase space thermodynamics of these metrics and their thermodynamic geometry.
	
	Working throughout in the extended phase space, we first rewrite the two-charge solution in a Schwarzschild-like coordinate chart and thus obtain compact closed forms for its thermodynamic quantities. We then study the equation of state, showing that the equal-charge case reduces to that of the Reissner--Nordstr\"om-AdS black hole, before turning to the distinct-charge case where the phase structure departs qualitatively from its Reissner--Nordstr\"om-AdS counterpart. We find that for sufficiently distinct charges a second critical point emerges, producing two separate (subcritical and supercritical) regions of small--large black hole coexistence at fixed pressure. We then construct the Ruppeiner geometry of the solution, taking the temperature and thermodynamic volume as fluctuation variables, and study its renormalized curvature. In the equal-charge limit we recover the known Reissner--Nordstr\"om-AdS results \cite{wei2019a, wei2019b}, properly excluding the regions of positive curvature that occur for de Sitter black holes with negative pressure and are excluded by  Maxwell's equal area law. In the distinct charge case, we show that the presence of the second critical point effectively `bends' the curve of vanishing Ruppeiner curvature towards lower volumes, so that both critical points occur in the region of attractive microscopic interactions.

	Our paper is organized as follows. In Section~\ref{stu} we present the STU black hole solution and its thermodynamic properties, focusing on the extended phase space and obtaining closed forms for the thermodynamic quantities in terms of the physical charges. In Section~\ref{eos} we examine the equation of state and study the resulting thermodynamic phase structure through the Gibbs free energy. In Section~\ref{ruppeiner}, we construct the associated Ruppeiner geometry and, through its renormalized curvature, probe the microscopic interactions of the solution. We summarize our results and comment on future extensions of this work in Section~\ref{conclusions}. Natural units where $\hbar=G=c=k_\text{B}=1$ are used throughout.

	\section{STU black holes and their thermodynamics}\label{stu}

	Broadly speaking, thermodynamic systems must possess a certain minimal number of independent parameters to exhibit rich thermodynamic behaviour and black holes are no exception. In anti-de Sitter space the addition of a $U(1)$ charge to the Schwarzschild solution brings with it new critical phenomena (the small-to-large black hole transition) \cite{chamblin1999a, chamblin1999b, cvetic1999a, kubiznak2012} that does not appear in the uncharged case \cite{hawking1983}. It is natural to ask what further structure emerges for black holes carrying several independent charges, and the STU model provides an ideal setting in which to test this. Prior to introducing the STU solution, we first briefly review the extended phase space black hole thermodynamics. 
	
	\subsection{Extended phase space thermodynamics}\label{lambda}

	For an asymptotically flat black hole, there is no quantity in the solution that naturally plays the role of a thermodynamic pressure, so the absence of a pressure-volume term from the first law is, on its face, unsurprising (see, however, the recent proposal \cite{borsboom2026}). In the presence of a cosmological constant, this picture changes, since $\Lambda$ itself acts like a fluid with pressure.
	
	It is then natural to ask how $\Lambda$ should enter the first law. The answer is to allow $\Lambda$ to vary, an idea explored at least as early as 1985 by Teitelboim \cite{teitelboim1985} and developed extensively since. Early treatments incorporated a pressure-volume term by evaluating the relevant quantities for specific spacetimes \cite{brown1994,creighton1995,caldarelli2000,wang2006}, while the general result is due to Kastor, Ray, and Traschen \cite{kastor2009}, whose scaling arguments and Hamiltonian perturbation analysis yield an {\it extended} first law without reference to any specific solution. In their work, a variation $\delta\Lambda$ in the cosmological constant enters the mass variation as
	\be\label{firstlaw3}
	\delta M=\dfrac{\kappa}{8\pi} \delta A+\dfrac{\Theta}{8\pi }\delta \Lambda+\dots\ 
	\ee
	Here $\Theta$ is a Komar integral of a particular Killing potential,  $\omega^{ab}$, and carries the interpretation of (minus) the volume that the black hole excludes from the spacetime. Identifying $\Lambda$ with  pressure, 
	$
	P\equiv-\Lambda/(8\pi),
	$
	the term $\Theta\,\delta\Lambda/(8\pi)$ becomes $V\,\delta P$. Eq.~\eqref{firstlaw3} then  assumes the form of an extended first law of thermodynamics,
	\be\label{firstlaw2}
	\delta M=T\delta S+V \delta P+\dots\ 
	\ee
	in which the thermodynamic volume 
	\be
	V \equiv\left(\frac{\partial M}{\partial P}\right)_{S, Q, J}
	\ee
	is the quantity conjugate to the pressure.
	
	An important distinction between this law and the first law without variable $\Lambda$ is that the right hand side now represents the variation of the enthalpy $H=U+PV$ of the system. We therefore identify the mass $M$ with $H$ rather than the internal energy $U$. The remainder of this work will take place in this extended phase space, where in general the cosmological constant in $d$ dimensions is related  to the thermodynamic pressure through
	\be
	P=-\frac{\Lambda}{8\pi}=\frac{g^2(d-1)(d-2)}{16\pi}\ ,
	\ee
	where $g$ is the inverse AdS length scale. 
	
	The phase behaviour of the system is most directly accessible through the  Gibbs free energy, which in the extended phase space is
	\be\label{gibbs}
	G\equiv M-TS=G(T,P,\dots)\ .
	\ee
	At fixed temperature and pressure, the equilibrium state of the system is the one that minimizes $G$. We examine $G$ for various STU black hole systems below.

	\subsection{The STU solution}

	The STU model appears in gauged $\mathcal{N}=8$, $SO(8)$ supergravity, which lives in the massless sector of $AdS_4\!\times\!\mathcal{S}^7$ M-theory. It couples gravity to three vector multiplets, whose complex scalar fields $S$, $T$, and $U$ parametrize the moduli space of the theory, and is invariant under a duality symmetry that permutes these fields (known as `triality' \cite{behrndt1996}). The gauging supplies a cosmological constant, so that the solutions of the theory are asymptotically anti-de Sitter. Within the $U(1)$ Cartan subgroup, the theory admits charged black hole solutions carrying anywhere from one to four distinct electric charges. These STU black holes generalize the familiar Einstein--Maxwell solutions of general relativity in a controlled way, reducing to them in appropriate limits while retaining some imprint of their string theory origin. The full bosonic part of the action within the $U(1)\!\times\!U(1)\!\times\!U(1)\!\times\!U(1)$ subgroup is
	\be\label{action}
	\mathcal{S}=\frac{1}{16 \pi}  \int d^{4} x \sqrt{-g}\Bigl(\mathcal{R}-\frac{1}{2} \sum_{i=1}^{3}\bigl(\partial_{\mu} \phi^{(i)}\bigr)^{2}-\mathcal{V}-\frac{1}{4} \sum_{I=1}^{4} X_{I}^{-2} F_{\mu \nu}^{(I) 2}\Bigr)\ ,
	\ee
	where writing $\vec{\phi}=(\phi^{(1)},\phi^{(2)},\phi^{(3)})$, we have
	\be
	X_I=e^{-\tfrac{1}{2}\vec{a}_I\cdot\vec{\phi}} \ ,\quad \vec{a}_1=(1,1,1)\ , \quad \vec{a}_2=(1,-1,-1)\ , \quad \vec{a}_3=(-1,1,-1)\ , \quad \vec{a}_4=(-1,-1,1)\ ,
	\ee
	with potential 
	\be
	\mathcal{V}=-g^2\sum_{I<J}X_I X_J\ ,
	\ee
	and  field strengths 
	\be
	F^{(I)}_{\mu\nu}=\partial_{\mu}A^{(I)}_{\nu}-\partial_{\nu}A^{(I)}_{\mu}\ .
	\ee
	This action represents supergravity coupled to three vector supermultiplets. The theory admits the following multi-charged black hole solution 
	\cite{behrndt1996}:
	\ba\label{metric}
	d s^{2}&=&-\left(H_{1} H_{2} H_{3} H_{4}\right)^{-1 / 2} {\tilde f}(\rho)\, d t^{2}+\left(H_{1} H_{2} H_{3} H_{4}\right)^{1 / 2}\Bigl(\frac{d \rho^{2}}{{\tilde f}(\rho)}+\rho^{2} d\Omega^{2}\Bigr)\ ,\nonumber\\
	A^{(I)}&=&\frac{\sqrt{q_{I}\left(q_{I}+2 m\right)}}{\rho+q_{I}} d t\,,\quad 
	X_{I}=\frac{\left(H_{1} H_{2} H_{3} H_{4}\right)^{1 / 4}}{H_{I}}\,,
	\ea
	where
	\be
	\tilde f(\rho)=1-\frac{2 m}{\rho}+g^{2} \rho^{2} H_{1} H_{2} H_{3} H_{4} \ ,\qquad H_{I}=1+\frac{q_{I}}{\rho}\ .
	\ee
	The solution is characterized by the following mass and physical charges:
	\be \label{charge}
	M=m+\frac{1}{4}\sum_I q_I\,,\quad Q_{I}=\frac{1}{2} \sqrt{q_{I}\left(q_{I}+2 m\right)}\,.
	\ee 
	In what follows we will specialize to the case where there are only two distinct physical charges,
	setting 
	\be \label{pair}
	q_3=q_1\,\quad \mbox{and}\quad q_4=q_2
	\ee 
	in the above. One can verify that the resulting expressions are the same as those obtained by considering only the $U(1)\!\times\! U(1)$ subgroup. This is distinct from the doubly charged case considered in \cite{caceres2015}, where two charges are taken to be equal and the rest are zero. 
	
	Note also that as written above the solution is not in Schwarzschild-like coordinates. This is most easily seen by considering the limit where all four charges are set equal ($q_I=q$). This should correspond to the Reissner--Nordstr\"om-AdS solution, but one finds instead that
	\be
	d s^{2}=-\left(1+\frac{q}{\rho}\right)^{-2} \tilde f(\rho)\, d t^{2}+\left(1+\frac{q}{\rho}\right)^{2}\left(\frac{d \rho^{2}}{\tilde f(\rho)}+\rho^{2} d \Omega^{2}\right)\ ,
	\ee
	with
	\be
	\tilde f(\rho)=1-\frac{2 m}{\rho}+g^{2}\left(1+\frac{q}{\rho}\right)^{4} \rho^{2}\ .
	\ee
	To obtain the familiar form of the Reissner--Nordstr\"om-AdS solution, one uses the fact that the ordinary Schwarzschild radial coordinate $r$ is related to $\rho$ through the transformation $\rho=r-q$. 
	
	\subsection{Schwarzschild-like form for pairwise equal charges}
	
	Let us first consider the pairwise-equal charge case \eqref{pair}. 
	Then, we have 
	\be\label{Hdef}
	H(\rho)\equiv\sqrt{H_1H_2H_3H_4}=\Big(1+\frac{q_1}{\rho}\Big)\Big(1+\frac{q_2}{\rho}\Big)\ ,
	\ee
	and the metric \eqref{metric} reduces to
	\be\label{metricreduced}
	ds^2=-\dfrac{\tilde f(\rho)}{H(\rho)}\,dt^2+\frac{H(\rho)}{\tilde f(\rho)}d\rho^2+H(\rho)\rho^2 d\Omega^2\,,\qquad 
	\tilde f(\rho)=1-\frac{2m}{\rho}+g^2\rho^2H(\rho)^2\ .
	\ee
	
\begin{figure}[h]
		\centering
		\includegraphics[width=0.43\textwidth]{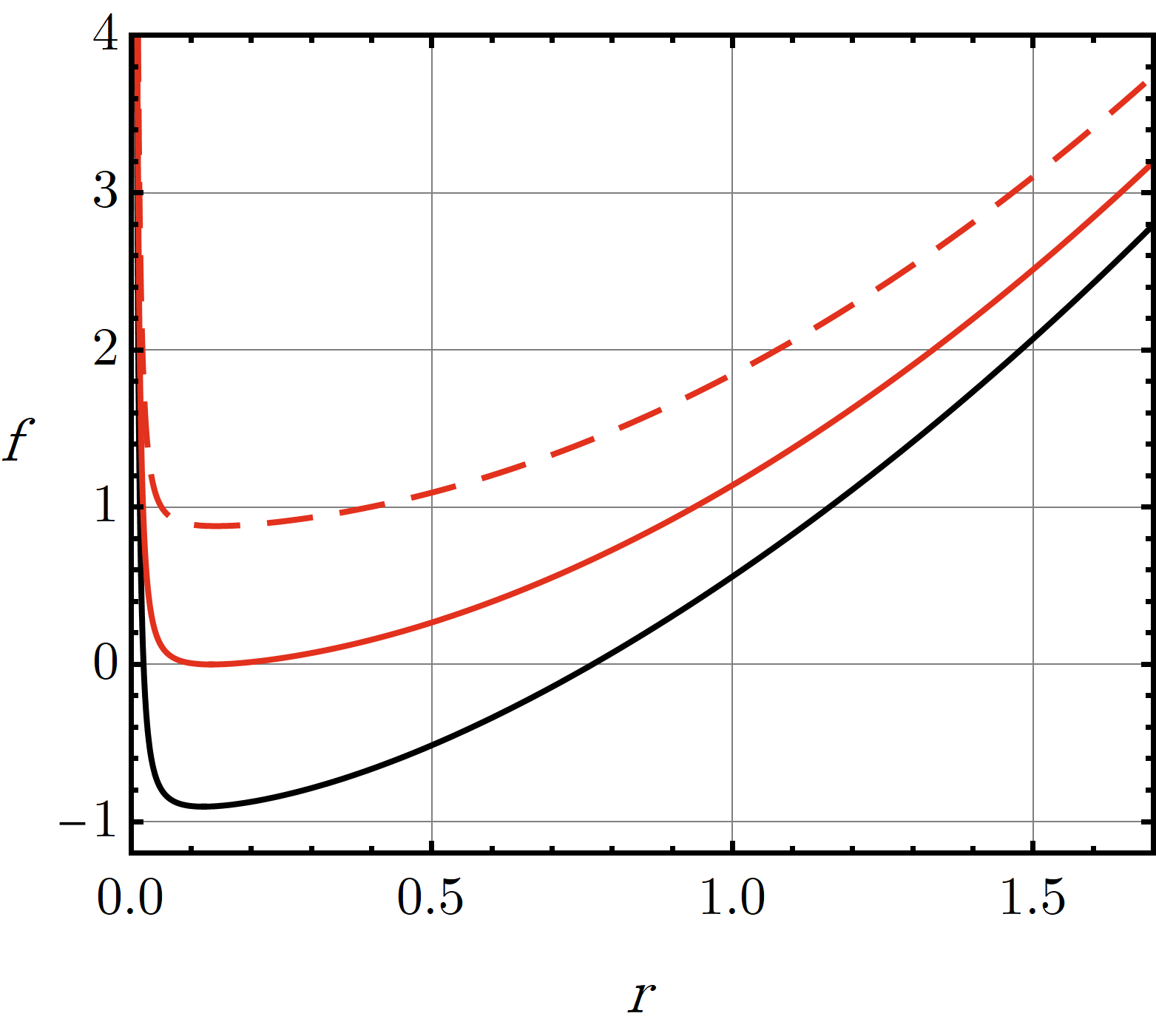}\qquad \includegraphics[width=0.43\textwidth]{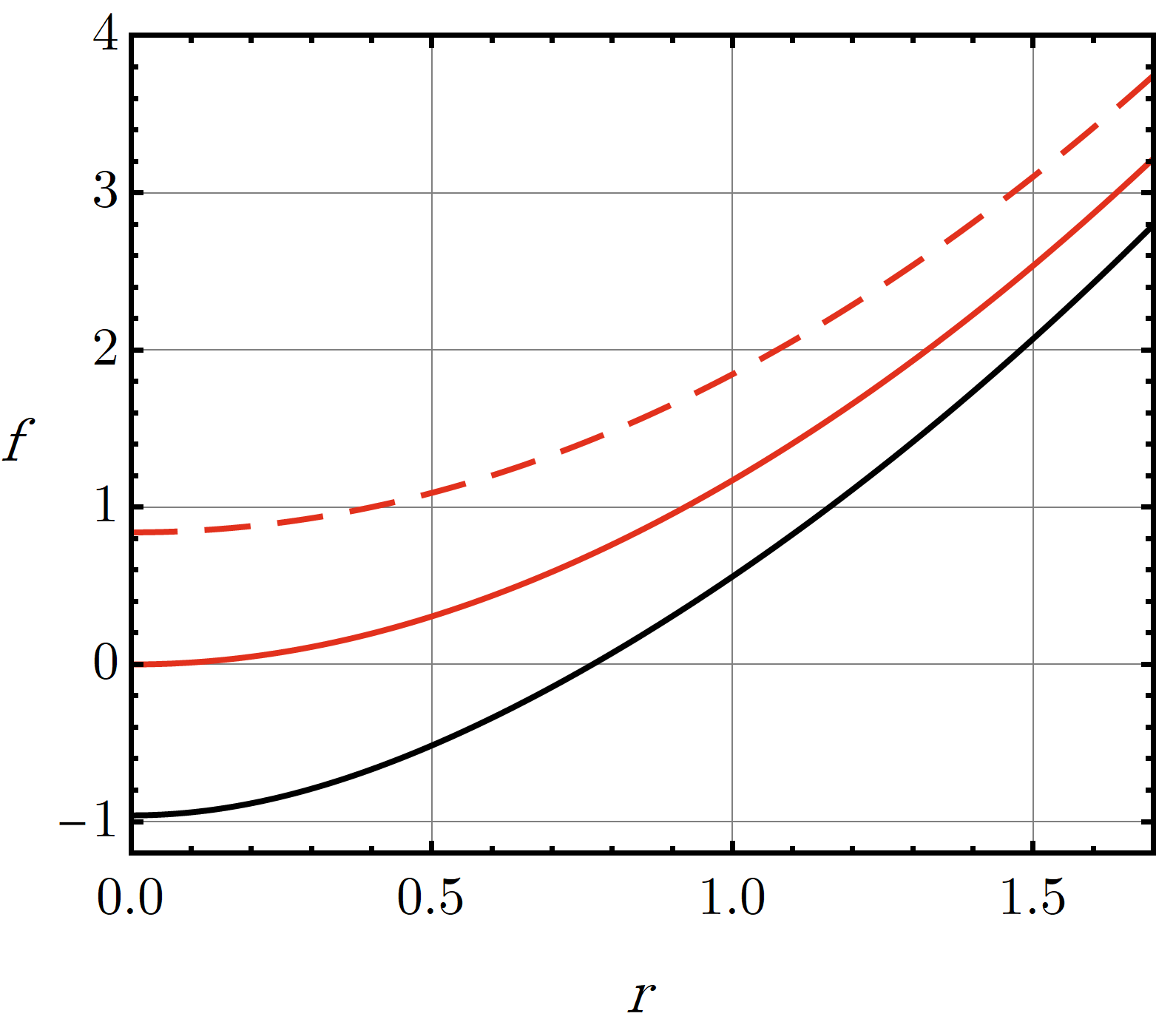}\hspace{0.6cm}
		\caption{{\bf Horizons of STU black holes.} We display the metric function $f$ of pairwise equal charge STU black holes; all quantities are scaled by $g$ (we set $g=1$ throughout). {\em Left:} The standard Reissner--Nordstr\"om-AdS-like behavior of STU holes that occurs when both charges are non-vanishing. As $M$ increases we observe a naked singularity (red dashed curve), an extremal black hole (red solid curve), and  a 2-horizon black hole (black curve). We have set $Q_1=1$ and $Q_2=0.01$. {\em Right:} Qualitatively distinct behavior when one of the charges vanishes (e.g. $Q_1=Q$ and $Q_2=0$). We observe again a naked singularity (red dashed curve), but also a vanishing radius hole for $M=Q$ (red solid curve), and single horizon black hole  (black curve) for $M>Q$. We have set $Q_1=1$ and $Q_2=0$.  
		}
		\label{fig0}
	\end{figure}

	To simplify our subsequent discussion, we introduce the areal radius and rewrite the metric in a Schwarzschild-like form in terms of physical charges. Namely, we demand the following  form of the metric,
	\be\label{schwarzmetric}
	ds^2=- f(r)\,dt^2+\frac{ h(r)}{f(r)}\,dr^2+r^2 d\Omega^2\ ,
	\ee
	defining the areal radius $r$ by
	\be\label{Rsq}
	r^2\equiv H(\rho)\,\rho^2=(\rho+q_1)(\rho+q_2)\quad \Leftrightarrow \quad 
	\rho=-\frac{q_1+q_2}{2}+\sqrt{\frac{(q_1-q_2)^2}{4}+r^2}\ .
	\ee
	In addition
	\ba 
	f(r)&\equiv& \frac{\tilde f(\rho)}{H(\rho)}=\frac{\rho-2m}{\rho H}+g^2\rho^2 H=\frac{\rho(\rho-2m)}{r^2}+g^2r^2\,\nonumber\\
	&=&
	1+\frac{2\left(Q_1^2+Q_2^2\right)}{r^2}-\frac{2}{r^2}\sqrt{\left(Q_1^2-Q_2^2\right)^2+M^2r^2}+g^2r^2\ ,
	\ea  
	where to write the last expression we have used the formulas for physical charges \eqref{charge}, together with 
	\be\label{qdiff}
	q_1-q_2=\frac{2\left(Q_1^2-Q_2^2\right)}{M}\ .
	\ee
	Finally, to have $H(\rho)\,\tilde f(\rho)^{-1}\,d\rho^2=h(r)\,f(r)^{-1}\,dr^2$, we have 
	\be\label{hdef}
	h(r)\equiv\left(\frac{d\rho}{dr}\right)^2=\frac{M^2r^2}{\left(Q_1^2-Q_2^2\right)^2+M^2r^2}\ .
	\ee
	The horizon is located at the largest root of 
	\be 
	f (r_+)=0\ .
	\ee 
	Moreover, we have the following expansion of $f$ for small $r$:
	\be 
	f=\frac{2\bigl[(Q_1^2+Q_2^2)-\bigl|Q_1^2-Q_2^2\bigr|\bigr]}{r^2}
	+1-\frac{M^2}{\bigl|Q_1^2-Q_2^2\bigr|}+O(r^2)\,.
	\ee 
	When both charges are distinct from zero, the first term diverges to plus infinity, and we observe the standard Reissner--Nordstr\"om-AdS-like behavior with two, one extremal, or zero horizons, as displayed in the left panel of Fig.~\ref{fig0}. However, more interestingly, when one of the charges vanishes, the first term vanishes and we recover a finite value for $f$ at the origin, namely 
	\be 
	f=\frac{Q^2-M^2}{Q^2}+O(r^2)\ ,
	\ee 
	where we have set $Q_1=Q$, $Q_2=0$ without loss of generality, although the Kretschmann scalar still diverges as $r\to 0$. It follows that there is a naked singularity for $M<Q$, a vanishing radius black hole for $M=Q$, and a single horizon black hole for $M>Q$, as displayed in the right panel of Fig.~\ref{fig0}. As we shall see, the two cases also have distinct thermodynamic behavior.    
	The thermodynamic quantities of pairwise-equal charge solutions are 
	\ba\label{TDs} 
	M&=&m+\frac{1}{2}(q_1+q_2)=
	\frac{\chi_+}{2r_+}\ ,\nonumber\\
	T&=&\frac{f'(r_+)}{4\pi \sqrt{h(r_+)}}=\frac{r_+^4+4r_+^4g^2(Q_1^2+Q_2^2)+4g^2r_+^6+3g^4r_+^8-16Q_1^2Q_2^2}{4\pi r_+^3\chi_+}\ ,\nonumber\\
	S&=&\pi r_+^2\ ,\\
	\phi_1&=&\frac{2Q_1}{r_+}\sqrt{\frac{r_+^2+4Q_2^2+r_+^4g^2}{r_+^2+4Q_1^2+r_+^4g^2}}\,,\quad 
	\phi_2=\frac{2Q_2}{r_+}\sqrt{\frac{r_+^2+4Q_1^2+r_+^4g^2}{r_+^2+4Q_2^2+r_+^4g^2}}\ ,\nonumber\\
	P&=&\frac{3g^2}{8\pi}\,,\quad 
	V=\left(\frac{\partial M}{\partial P}\right)_{\!\!S,Q_1,Q_2}=\frac{4}{3}\frac{\pi r_+^3}{\chi_+}\left(r_+^2+2\left(Q_1^2+Q_2^2\right)+g^2r_+^4\right)\ ,\nonumber
	\ea 
	where $\phi_1$ and $\phi_2$ are the respective chemical potentials, and we have denoted
	\be\label{chiplus}
	\chi_+\equiv\sqrt{\left(r_+^2+4Q_1^2+g^2r_+^4\right)\left(r_+^2+4Q_2^2+g^2r_+^4\right)}\ .
	\ee
	It is easy to verify that these quantities satisfy the corresponding (extended) first law and Smarr relations:
	\ba 
	\delta M&=&T\delta S+\phi_1 \delta Q_1+\phi_2 \delta Q_2+V\delta P\ ,\\
	M&=&2TS+\phi_1 Q_1+\phi_2 Q_2-2PV\,.
	\ea

	\section{The equation of state and free energy}\label{eos}
	
	Having written down the above explicit expressions for thermodynamic quantities, we now proceed to study the corresponding equation of state
	\be \label{EoS}
	P=P(V,T,Q_1,Q_2)\quad \Leftrightarrow \quad T=T(P,V,Q_1,Q_2)\ ,
	\ee 
	where the latter is implicitly given by second relation in \eqref{TDs}, and the behavior of the (canonical ensemble) Gibbs free energy
	\ba \label{Fdef}
	G&=&M-TS=G(T,P, Q_1,Q_2)\nonumber\\
	&=&
	\frac{r_+^4+4r_+^2(r_+^2g^2+2)(Q_1^2+Q_2^2)-g^4r_+^8+48Q_1^2Q_2^2}{4 r_+\chi_+}\ ,
	\ea 
	where $\chi_+$ is given in \eqref{chiplus}.

	We consider 3 cases: i) the Reissner--Nordstr\"om-AdS black hole where all charges are equal ($Q=Q_1=Q_2$) ii) the `single charge' case where $Q=Q_1$, $Q_2=0$ and iii) the distinct `two-charge' case where $0\neq Q_1\neq Q_2\neq 0$, extending thus the studies in \cite{kubiznak2012,caceres2015,sekhmani2023}. 
	\subsection{Reissner--Nordstr\"om-AdS}\label{rnads}
\begin{figure}[h]
		\centering
		\includegraphics[width=0.45\textwidth]{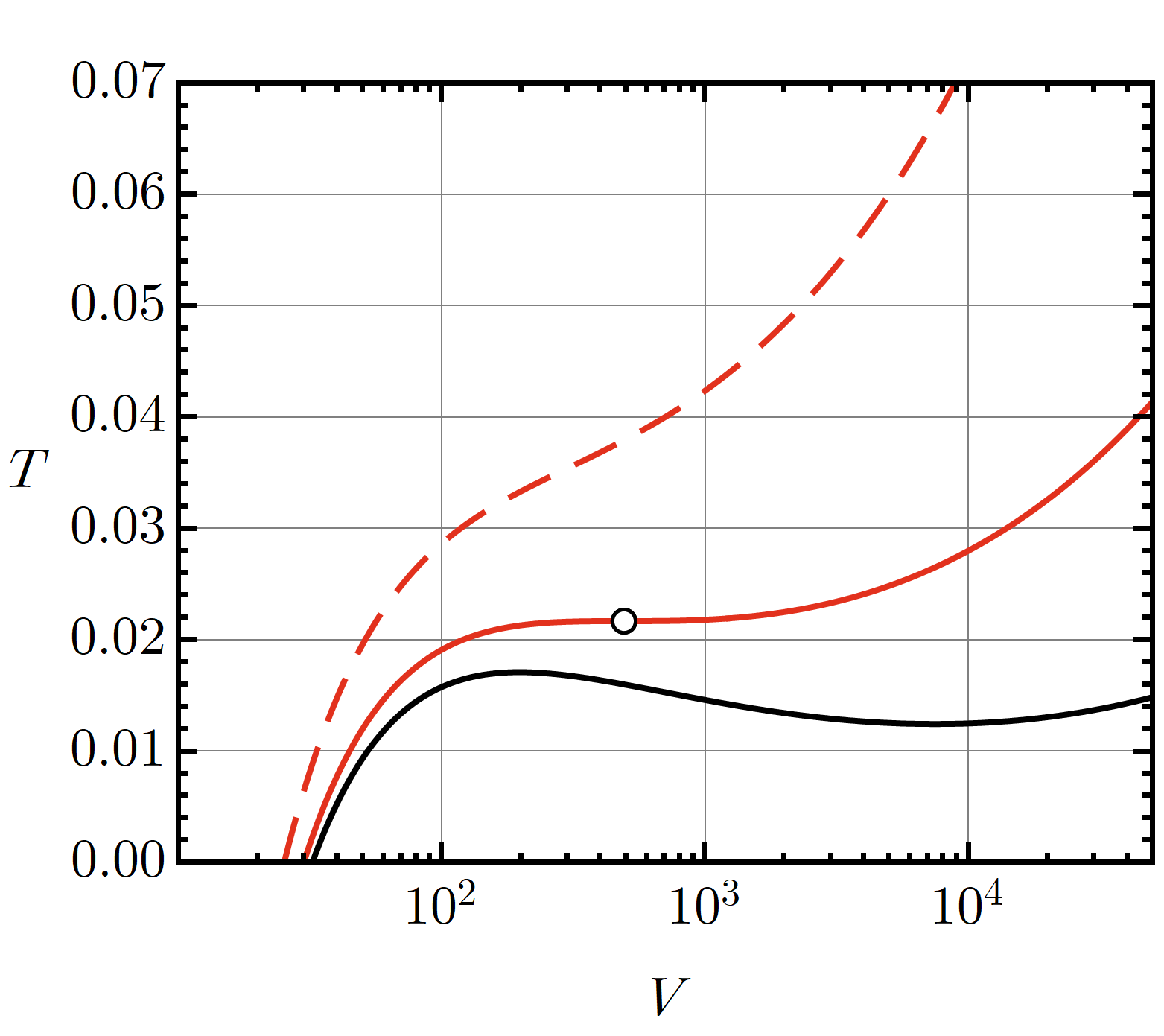}\qquad \includegraphics[width=0.45\textwidth]{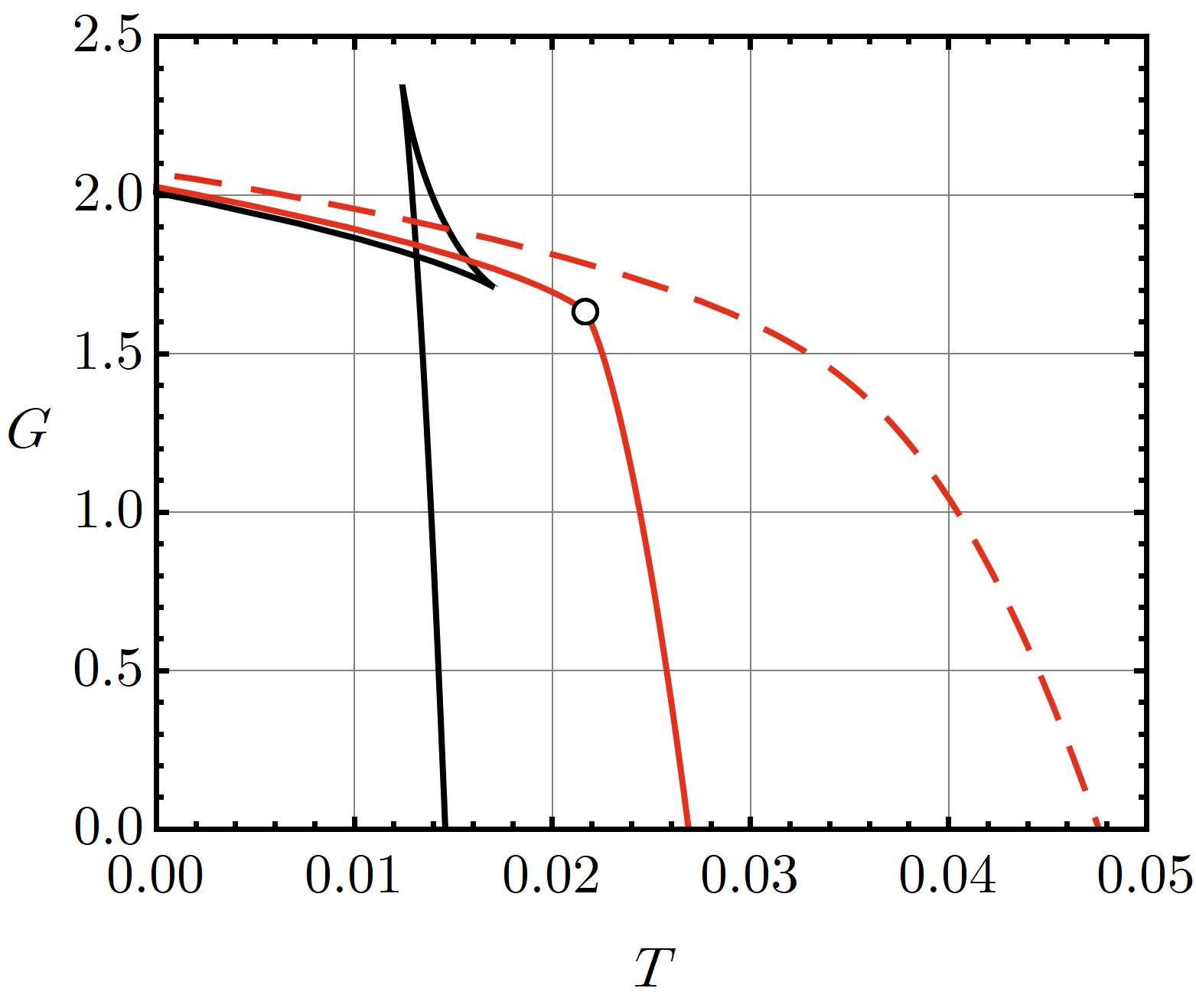}\hspace{0.5cm}
		\caption{{\bf Reissner--Nordstr\"om-AdS case.} We display the standard $T-V$ diagram (left) and the $G-T$ diagram (right) with $Q=1$. The small black circle indicates the critical point. It lies on the $P=P_c$ isobar (indicated by the solid red curve) in both diagrams. The isobars on the left are plotted for $P/P_c=\{0, 0.3, 1, 3\}$ from bottom up, and similarly on the right with the $P=0$ isobar absent.} 
		\label{fig1}
	\end{figure}

	When all charges are equal, 
	\be 
	Q\equiv Q_1=Q_2\ ,
	\ee 
	the metric reduces to that of the ordinary Reissner--Nordstr\"om-AdS black hole. Likewise, the relevant thermodynamic quantities \eqref{TDs} also reduce  to their Reissner--Nordstr\"om-AdS forms, studied previously in \cite{kubiznak2012}. Explicitly, we have
	\ba\label{TDsRN}
	M&=&\frac{r_+^2+4Q^2+g^2r_+^4}{2r_+}\,,\quad 
	T =\frac{3g^{2}r_+^{4}+r_+^{2}-4Q^{2}}{4\pi r_+^{3}}\,,\quad  
	S =\pi r_+^{2}\,, \nonumber\\
	\phi&\equiv&\phi_1=\phi_2=\frac{2Q}{r_+}\,, \quad P=\frac{3g^2}{8\pi}\,,\quad 
	V =\frac{4}{3}\pi r_+^{3}\,,
	\ea
	and the free energy reads 
	\be 
	G(T,P,Q)=M-TS=\frac{r_+^2+12Q^2-g^2r_+^4}{4r_+}\ .
	\ee 
	Using expressions for $P$ and $V$, the second relation in \eqref{TDsRN} immediately yields the following equation of state \eqref{EoS}: 
	\be
	T(V,P,Q)=P\biggl(\frac{6V}{\pi}\biggr)^{1/3}+\frac{1}{12}\biggl(\frac{36}{\pi^2V}\biggr)^{1/3}-\frac{4Q^2}{3V}\,,
	\ee
	and its inversion:
	\be 
	P(T,V,Q)=T\biggl(\frac{\pi}{6V}\biggr)^{1/3}+\frac{2Q^2}{9}\biggl(\frac{36\pi}{V^4}\biggr)^{1/3}-\frac{1}{12}\biggl(\frac{6}{\pi V^2}\biggr)^{1/3}\,,
	\ee 
	which is identical to the equation of state obtained in \cite{kubiznak2012} upon rescaling the current formulas as $2Q\to Q$, accounting for 4 identical $U(1)$ charges. We reproduce the corresponding $T-V$ and $G-T$ diagrams in Fig.~\ref{fig1},   with the latter plotted parametrically using $r_+$ as the corresponding parameter.

	As is well known 
	\cite{kubiznak2012,caceres2015}, there is a critical point, obtained from 
	\be\label{crit}
	\left(\dfrac{\partial T}{\partial V}\right)_{\!P=P_c}=\left(\dfrac{\partial^2 T}{\partial V^2}\right)_{\!P=P_c}=0\,,
	\ee
	giving the corresponding critical quantities:
	\be\label{rncrit}
	P_{c}=\frac{1}{384 \pi Q^{2}}\ , \quad V_{c}=64 \sqrt{6} \pi Q^{3},\  \quad  T_{c}=\dfrac{1}{6 \sqrt{6} \pi Q}\,.
	\ee
	Below $(T_c, P_c)$, we observe Van der Waals-like oscillations in the $T-V$ (and also $P-V$)  diagrams, corresponding to a first-order phase transition from small to large black holes, derivable from Maxwell's equal area law. Above these critical values, one observes a monotonic behavior of the isobars in the $T-V$ diagram, see left Fig.~\ref{fig1}, or,  alternatively, a monotonic behavior of the isotherms in the $P-V$ diagram. This is reflected by the behavior of the Gibbs free energy displayed in right Fig.~\ref{fig1} where we observe the characteristic swallowtail for $P<P_c$ indicating the first order phase transition, whereas the free energy is smooth for $P>P_c$.\footnote{In more detail, below the critical pressure the Gibbs free energy features the characteristic swallowtail, whose self-intersection marks a first-order transition between  small and large black holes. The near-horizontal branch corresponds to the small black hole and the near-vertical branch to the large one. The horizon radius (the order parameter) increases monotonically along the curve, and is thus discontinuous at the crossing as the system passes from one branch to the other. As the pressure is raised the swallowtail shrinks, closing to a point at the critical pressure, above which no transition occurs, though it is possible to obtain a Widom line in the supercritical region that thermodynamically distinguishes small-type black holes from large ones \cite{xu2026}. This is the same structure seen in previous studies of the Reissner--Nordstr\"om-AdS black hole \cite{kubiznak2012}.}

	\subsection{Single charge}
	
	\begin{figure}[h]
		\centering    \includegraphics[width=0.45\textwidth]{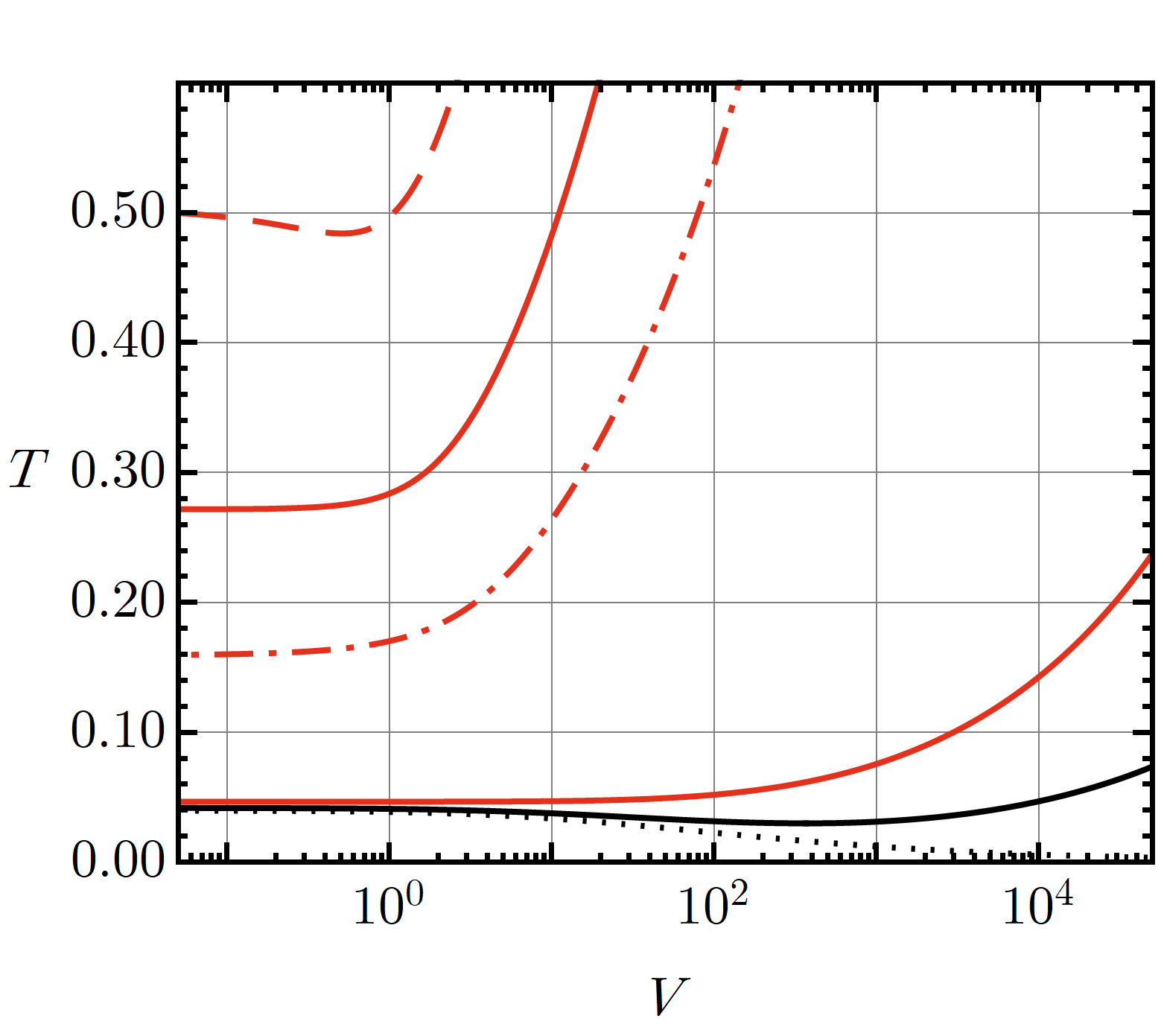}\qquad \includegraphics[width=0.45\textwidth]{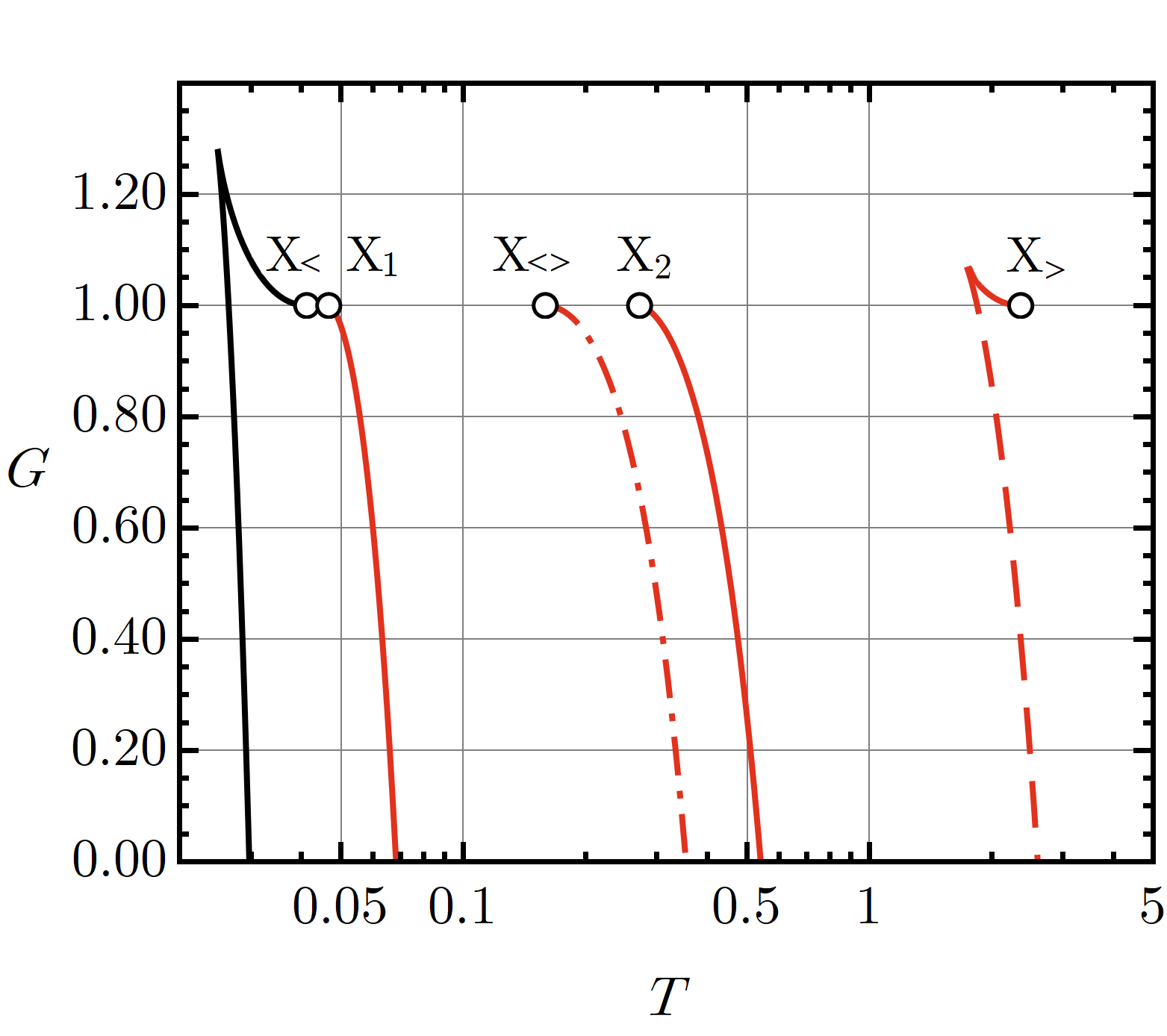}\hspace{0.5cm}
		\caption{{\bf Single charge case.} We display the standard $T-V$ diagram (left) and the $G-T$ diagram (right) with $Q=1$. The solid red curves indicate the `critical pressures' $P_{1}$ and $P_{2}$. Isobars on the left are plotted for $P=\{0.3 P_1, P_1,(P_1+P_2)/2, P_2, 2P_2\}$ from bottom up. Isobars on the right correspond to $P=\{0.2P_1, P_1, (P_1+P_2)/2, P_2, 10 P_2\}$ from left to right. Points $X$ are indicated by small black circles and from left to right denoted by $\{X_<, X_1, X_{<>}, X_2, X_{>}\}$.     }\label{fig2}
	\end{figure}
	
	Let us next consider the situation of a `single charge', namely
	\be 
	Q_1=Q\,,\quad Q_2=0\ .
	\ee 
	With this, the thermodynamic quantities \eqref{TDs} simplify as follows:
	\ba\label{singleTDs} 
	M&=&\frac{1}{2}\sqrt{(r_+^2g^2+1)(r_+^4g^2+4Q^2+r_+^2)}\ ,\nonumber\\
	T&=&\frac{3g^4r_+^4+4Q^2g^2+4r_+^2g^2+1}{4\pi \sqrt{(r_+^2g^2+1)(r_+^4g^2+4Q^2+r_+^2)}}\,,\quad S=\pi r_+^2\ ,\nonumber\\
	\phi&\equiv&\phi_1=2Q\sqrt{\frac{1+r_+^2g^2}{r_+^2+4Q^2+r_+^4g^2}}\,,\quad \phi_2=0\ ,\nonumber\\
	P&=&\frac{3g^2}{8\pi}\,,\quad 
	V=\frac{4}{3}\pi r_+^2\frac{r_+^4g^2+2Q^2+r_+^2}{\sqrt{(1+r_+^2g^2)(r_+^2+4Q^2+r_+^4g^2)}}
	\ea 
	and $G$ reads 
	\be 
	G=\frac{r_+^2+8Q^2+4r_+^2g^2Q^2-r_+^6g^4}{4\sqrt{(1+r_+^2g^2)(r_+^2+4Q^2+r_+^4g^2)}}\ .
	\ee 
	Since the volume $V$ is no longer a simple function of the horizon radius, both the equation of state $T(V,P,Q)$ and the Gibbs free energy $G(T,P,Q)$ have to be plotted parametrically. They are displayed for $Q=1$ in Fig.~\ref{fig2}.

	In this case, there are no critical points. However, there exists an `initial point' $X$,  at which the free energy originates, where the size of the hole vanishes ($r_+\to 0$). This point is characterized by 
	\be 
	T_X=\frac{1}{8\pi |Q|}\Bigl(1+\frac{32\pi }{3}PQ^2\Bigr)\,,\quad G_X=|Q|\ .
	\ee 
	As the pressure increases, the point $X$ moves to higher temperatures. In addition, there are now two `critical points', corresponding to two special positions of the $X$ points, given by:
	\be 
	X_1:\quad P_1=\frac{3}{32\pi Q^2}(3-2\sqrt{2})\,,\quad V_1=0\,,\quad T_1= 
	\frac{(2-\sqrt{2})}{4\pi |Q|}\,,
	\ee 
	and 
	\be 
	X_2:\quad 
	P_2=\frac{3}{32\pi Q^2}(3+2\sqrt{2})\,,\quad V_2=0\,,\quad T_2= 
	\frac{(2+\sqrt{2})}{4\pi |Q|}\ .
	\ee 
	For pressures between these critical pressures, $P\in (P_{1},P_{2})$, the free energy curve originates from the $X$ point and smoothly evolves to higher temperatures. This is the behavior which for the two charge case will lead to a `critical region' without phase transitions, as discussed in the next subsection. For  $P<P_{1}$, as $r_+$ is increased from zero, the free energy first moves towards smaller temperatures, reaches a cusp, and then smoothly moves towards high temperatures, as shown by the leftmost curve in the right panel of Fig.~\ref{fig2}. This is the case that will give rise to a (subcritical) standard swallowtail and a first order phase transition in the two charge case (with $P_{1}$ replaced by the corresponding  pressure $P_{c_1}$ of the critical point dependent on the second charge). Finally, for $P>P_{2}$, we again develop a cusp (the rightmost curve in the right panel of Fig.~\ref{fig2}.), explaining the origin of the second critical point in the two-charge case and the corresponding `reverse swallowtail' in the supercritical region.

	\subsection{Two unequal charges}\label{unequal}
	
	The case with unequal charges exhibits markedly different behavior from the previous two cases. The thermodynamic quantities are now given by the full expressions in \eqref{TDs}, and the behavior crucially depends on the ratio of the two charges.

	\begin{figure}[h]
		\centering    \includegraphics[width=0.45\textwidth]{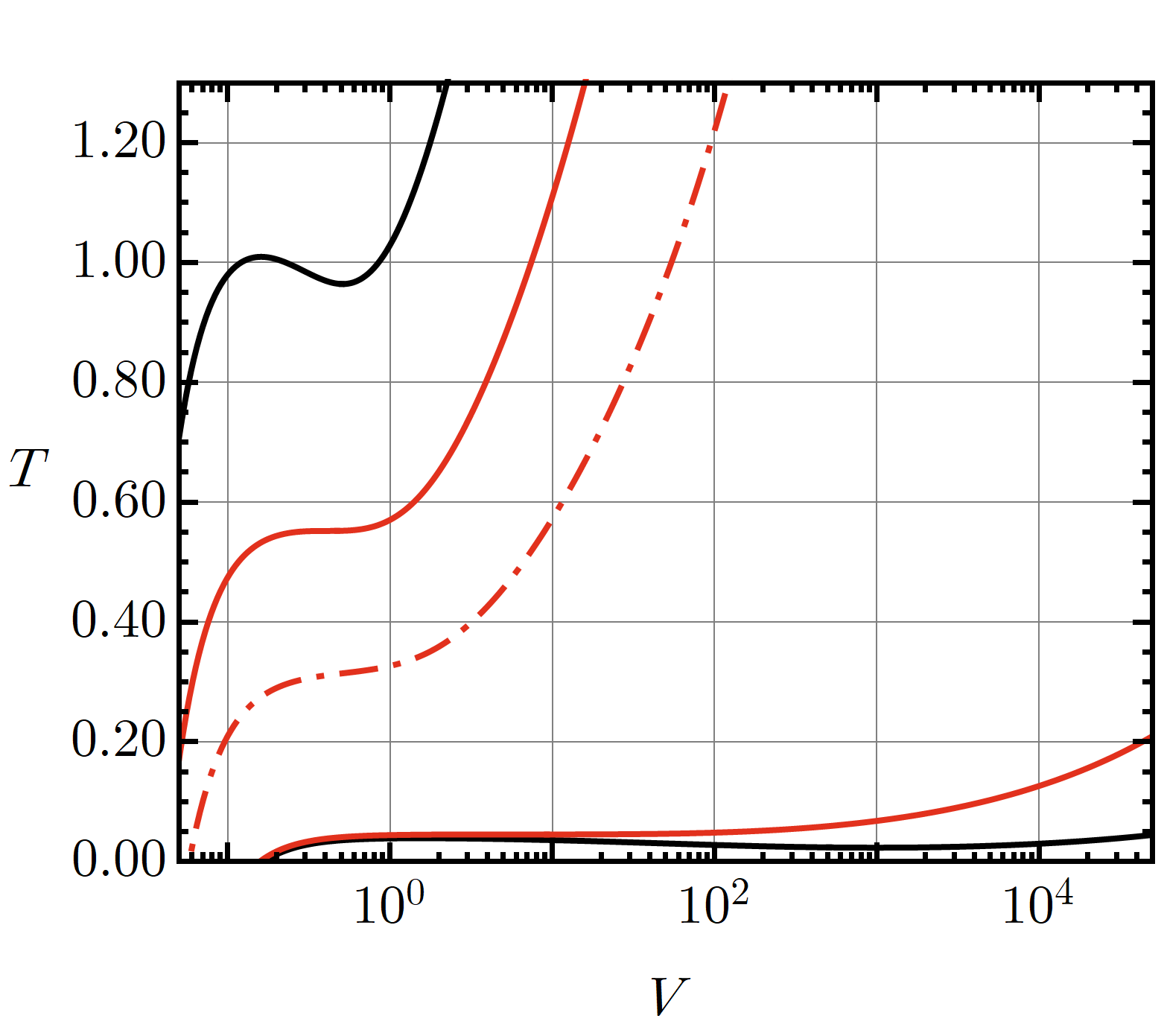}\qquad \includegraphics[width=0.45\textwidth]{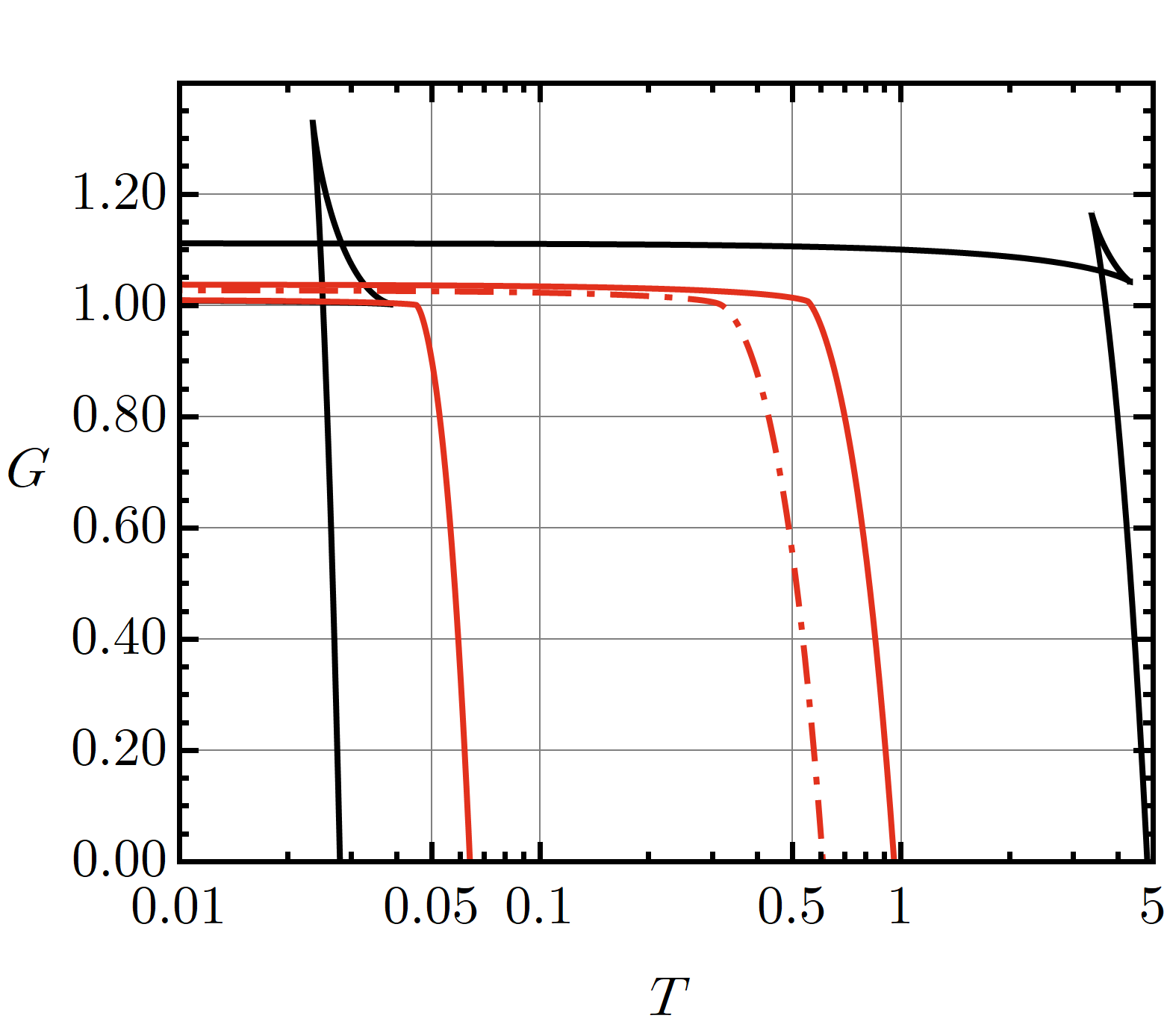}\hspace{0.5cm}
		\caption{{\bf Two  charge case.} We display the standard $T-V$ diagram and the $G-T$ diagram in the regime where there are two critical points, with $\epsilon=0.01$ and $Q=1$. {\it Left:} The $T-V$ diagram is plotted for $P=\{0, 0.2P_{c_1}, P_{c_1}, (P_{c_1}+P_{c_2})/2, P_{c_2}, 2 P_{c_2}\}$  from bottom up. {\it Right:} The $G-T$ diagram is displayed for  $P=\{0.2P_{c_1}, P_{c_1}, (P_{c_1}+P_{c_2})/2, P_{c_2}, 10 P_{c_2}\}$.  The solid red curves correspond to $P_{c_1}$ and $P_{c_2}$.}\label{fig3}
	\end{figure}

	\begin{figure}[h]
		\centering    \includegraphics[width=0.85\textwidth, height=8cm]{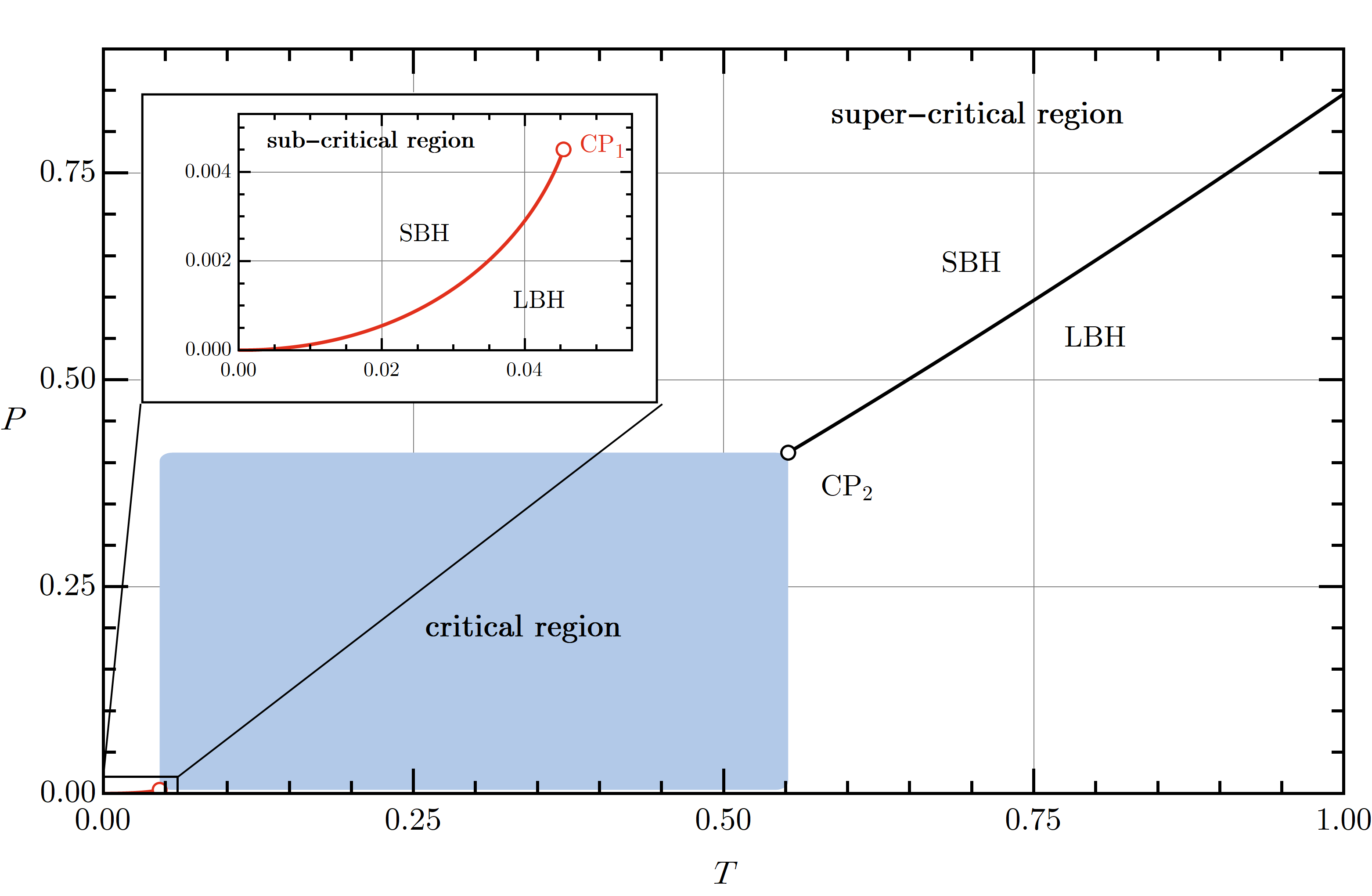}
		\caption{{\bf Two  charge case $P-T$ phase diagram.} The $P-T$ diagram is displayed for $\epsilon=0.01$ and $Q=1$. We observe 3 regions: a {\em sub-critical} region displaying the standard Van der Waals-like behavior is located in the bottom left and is magnified in the inset. The {\em Critical region} lies `between' the two critical points $(\mathrm{CP}_1)$ and $(\mathrm{CP}_2)$ and is highlighted in blue. In this region there is no distinction between small black holes (SBH) and large black holes (LBH). The {\em super-critical region} corresponds to the `reverse Van der Waals' and is displayed above the second critical point ($\mathrm{CP}_2$).  }   \label{figPT}
	\end{figure}
	
	\begin{figure}[h]
		\centering    \includegraphics[width=0.7\textwidth]{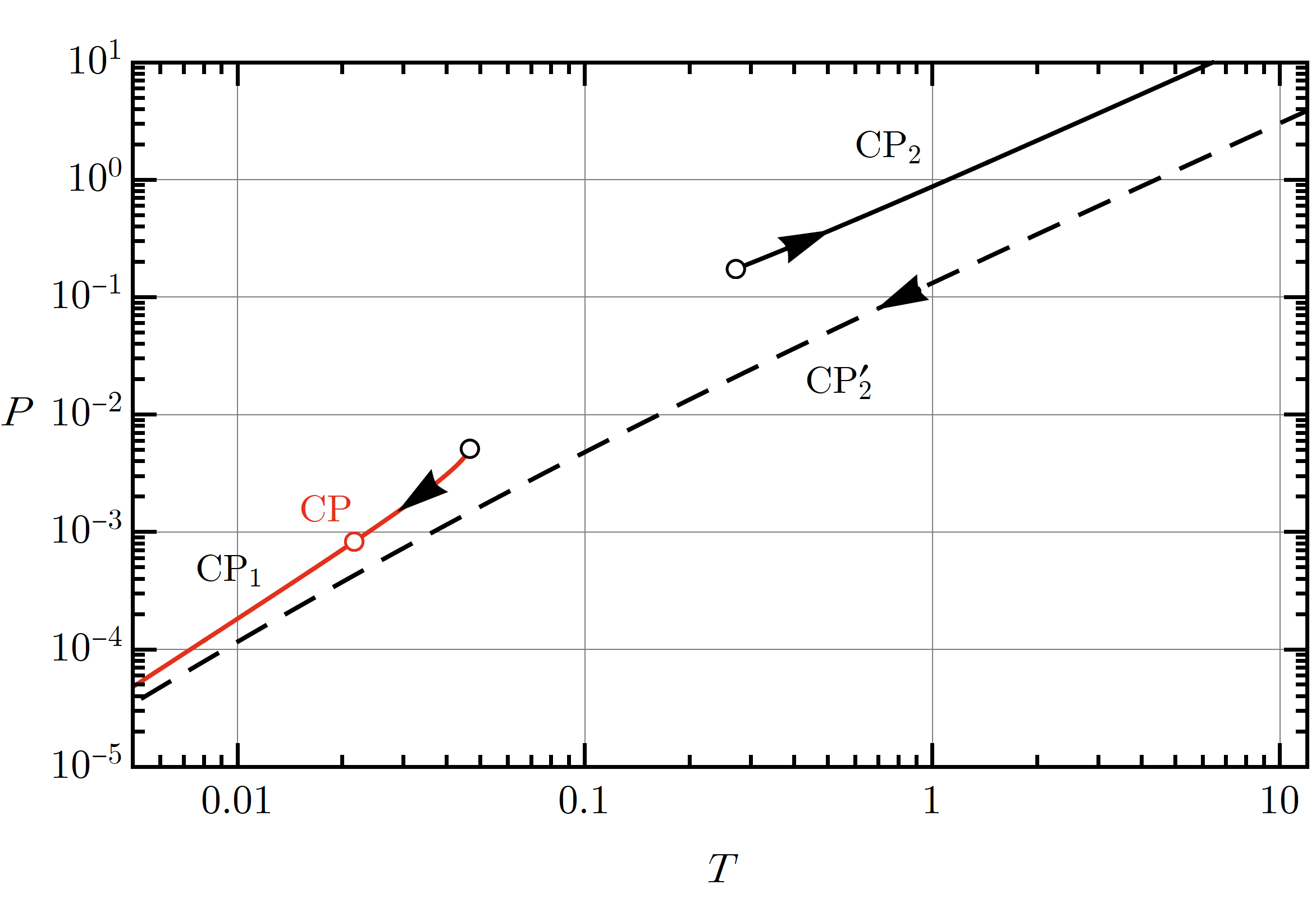}\hspace{1cm}
		\caption{{\bf Two  charge case: $\epsilon$-dependence of critical points.} `Motion' of critical points is displayed as $\epsilon$ increases (indicated by arrows) from $10^{-14}$ to $\epsilon=21$ and $Q=1$. As $\epsilon$ increases from zero, the left critical point $\mathrm{CP}_1$ moves towards lower pressures and temperatures (red curve). At the same time, the second  critical point $(\mathrm{CP}_2)$ moves towards higher pressures and temperatures, reaching infinite values at $\epsilon=\epsilon_c$ after which it disappears. For  $\epsilon\in (\epsilon_c,1/\epsilon_c)$ only the first critical point ($\mathrm{CP}_1$) exists and keeps receding towards smaller pressures and temperatures, in particular passing through the equal charge critical point $(\mathrm{CP})$ at $\epsilon=1$ indicated by a small red circle. As $\epsilon$ is increased above $1/\epsilon_c$, the second  (dashed black) branch of critical points (denoted here as $\mathrm{CP}_2'$) re-emerges from infinity and recedes towards lower pressures and temperatures. For $\epsilon>1/\epsilon_c$, we thus again observe two  critical points, denoted here by  $\mathrm{CP}_1$ and $\mathrm{CP}_2'$.}  \label{fig4}
	\end{figure}

	When a small additional charge $Q_2$ is added to the single charged case, that is for small charge ratios, 
	\be 
	\epsilon\equiv \frac{Q_2}{Q_1}\ll 1\ ,
	\ee 
	we observe two critical points, visible as inflection points in the $T-V$ curve and crossings in the $G-T$ curve of Fig.~\ref{fig3}. Their origin can can be intuitively understood in a way similar to what happens when a standard Maxwell charge is added to the Schwarzschild-AdS solution, e.g. \cite{altamirano2014b}. Namely, adding the additional charge $Q_2$ to the single-charged case introduces an additional small black hole branch that  originates from the $T=0$ extremal black hole and `joins' what used to be the $X$ point. This results in two critical points (CP$_1$ and CP$_2$) characterized by $(P_{c_1},V_{c_1}, T_{c_1})$ and $(P_{c_2},V_{c_2}, T_{c_2})$, which, although given by \eqref{crit}, cannot be determined analytically. In what follows, we assume that $P_{c_1}<P_{c_2}$ without loss of generality, as the two points can never coincide (there is no analogue of an isolated critical point \cite{dolan2014,  hu2024} for STU black holes with any $Q_2/Q_1$ ratio), and display the corresponding $P-T$ phase diagram for $\epsilon=0.01$ and $Q=1$ in Fig.~\ref{figPT}.

	For $P<P_{c_1}$, we observe the standard swallowtail and the first-order phase transition that terminates at the first critical point CP$_1$. We call this region {\it sub-critical}, enlarged in the inset in Fig.~\ref{figPT}. For pressures between $P_{c_1}$ and $P_{c_2}$, we observe smooth free energy -- this corresponds to the `{\em critical region}' where no sharp distinction between small and large black holes can be made.   Finally, for $P>P_{c_2}$, there is yet another `{\em reverse swallowtail}', with the corresponding first order phase transition `reversed'\footnote{The reversed phase transition is  absent below the critical pressure and present above it. It has previously been observed, for example, in higher-dimensional Lovelock black holes with hyperbolic horizons \cite{frassino2014}. Aside from its `pressure inversion' it shares the features of the ordinary swallowtail.} and emanating from the second critical point CP$_2$. We call the corresponding region `{\it super-critical}', see Fig.~\ref{figPT}. The second critical point introduces a further pair of small and large black hole phases at high pressure, which we term `super-critical', separated from the original `sub-critical' transition by the `critical phase'. This gives a nominal count of five phases: small and large sub-critical black holes, critical black holes, and  small and large super-critical phases. Because the sub- and super-critical regions are not separated by any genuine phase boundary, the only true coexistence curves being the two that terminate at the critical points, the system may pass between them without undergoing any phase transitions by a suitable variation of temperature and pressure. We note also that, because the critical points are not available in closed form, the critical exponents cannot be determined analytically.

	As the charge ratio $\epsilon$ increases from zero, the first critical point CP$_1$ moves towards lower pressures and temperatures, whereas the second critical point CP$_2$ moves towards larger and larger temperatures and pressures, as displayed in Fig.~\ref{fig4} (with $\mathrm{CP}_1$ denoted by a solid red curve and $\mathrm{CP}_2$ by a black one). Interestingly, as we approach 
	\be 
	\epsilon=\epsilon_c\equiv  {7-4\sqrt{3}}\approx 0.0717967697\ ,
	\ee 
	(see Appendix~\ref{A3} for derivation of this bound) the second critical point moves to infinity, with both $P_{c_{2}}$ and $T_{c_2}$ diverging and $r_{c_2}$ vanishing. Slightly above $\epsilon_c$, 
	the second critical point disappears and only the first critical point survives. Namely, for `comparable charge ratios' given by 
	\be 
	\epsilon\in (\epsilon_c, 1/\epsilon_c)\,,
	\ee  
	we observe a behavior qualitatively similar to the Reissner--Nordstr\"om case. In particular, as the charge ratio approaches unity, the first critical point coincides with the equal charge case, \eqref{rncrit}, denoted by CP in Fig.~\ref{fig4}. 
	
	At $\epsilon=1/\epsilon_c = 7+4\sqrt{3}$ the second critical point  (denoted by $\mathrm{CP}_2'$ and dashed black curve in Fig.~\ref{fig4}) re-emerges `from infinity', and as $\epsilon$ increases, it moves along a new branch towards lower pressures and temperatures. At the same time the first critical point $\mathrm{CP}_1$ remains present and  keeps receding towards smaller pressures and temperatures. Thus, for $\epsilon>1/\epsilon_c$ we again observe two critical points, now denoted by $\mathrm{CP}_1$ and $\mathrm{CP}_2'$ in the figure.\footnote{Note that by setting $Q=1$ in this graph, we have broken the $Q_1\leftrightarrow Q_2$ symmetry.} 
	
	To summarize, in the presence of both charges the phase behavior is quite involved and varies from the presence of a single standard critical point for $\epsilon\in(\epsilon_c, 1/\epsilon_c)$, to the presence of two critical points for $\epsilon<\epsilon_c$ or $\epsilon>1/\epsilon_c$. We shall now discuss how the potential  presence of the second critical point affects the Ruppeiner curvature.

	\section{Ruppeiner Geometry}\label{ruppeiner}
	
	\subsection{Ruppeiner curvature and black hole microstructure}

	We now turn to the application of the Ruppeiner geometry to the STU black hole. The general philosophy underpinning geometric thermodynamics is to equip the space of equilibrium states of a system with a Riemannian metric built from the second derivatives of some thermodynamic potential. One standard choice is the Weinhold metric, which is the Hessian of the internal energy with respect to the extensive variables, $g^{W}_{\mu\nu}=\partial_\mu\partial_\nu U$, with $U=M-PV$ related to the enthalpy $M$ of Section~\ref{stu} by a Legendre transform. The other, which we consider here, is the Ruppeiner metric \cite{ruppeiner1995}. It is (minus) the Hessian of the entropy, 
	\be \label{gR}
	g^{R}_{\mu\nu}=-\partial_\mu\partial_\nu S\ .
	\ee 
	It is conformally related to the Weinhold metric \cite{ruppeiner1995} through $g^{R}=g^{W}/T$, with the temperature $T$ as the conformal factor, so that either may be used to describe the same geometry. We work with the Ruppeiner metric because it has a direct statistical meaning. In the Gaussian theory of equilibrium fluctuations, the probability of a fluctuation between two nearby states is governed by the line element, with a larger separation corresponding to a less likely fluctuation \cite{wei2019b}. The curvature of this metric, in turn, reflects the correlations among the underlying degrees of freedom.

	In principle, one may study the system in response to fluctuations in any number of its extensive variables, giving a metric of the corresponding dimension. In practice, one restricts to a two-dimensional subspace spanned by a single pair of fluctuation variables, which keeps the geometry simple. The natural pair is the temperature $T$ (conjugate to the entropy), together with one further extensive quantity $x\in\{V,Q,J,\dots\}$. Passing to these variables amounts to a Legendre transform from the energy to the free energy $F$,
	\be 
	F=U-TS\ ,
	\ee 
	in terms of which the entropy Hessian  \eqref{gR} 
	takes the form
	\be\label{ruppeiner0}
	g_{\mu \nu}=\frac{1}{T}\left(\begin{array}{cc}
		-\left(\frac{\partial^{2} F}{\partial T^{2}}\right)_{x} & 0 \\
		0 & \left(\frac{\partial^{2} F}{\partial x^{2}}\right)_{T}
	\end{array}\right)=\frac{1}{T}\left(\begin{array}{cc}
		\left(\frac{\partial S}{\partial T}\right)_{x} & 0 \\
		0 & -\left(\frac{\partial y}{\partial x}\right)_{T}
	\end{array}\right)\ ,
	\ee
	where $y$ denotes the intensive variable conjugate to $x$, $y\in \{P,\phi, \Omega,\dots\}$. In what follows, we take the second fluctuation variable to be the thermodynamic volume, $x=V$, implying $y=P$. This is the natural choice here since the charges are held fixed in the canonical ensemble and the solutions are non-rotating, so the volume is the remaining extensive variable, and the resulting curvature probes the same small--large black hole structure as the equation of state in Section~\ref{eos}.
	
	It is the first law that reduces \eqref{ruppeiner0} to a form fixed by the equation of state. The relevant potential is the free energy $F$, whose differential is given by the first law as $dF=-S\,dT-P\,dV$, so that $S=-(\partial F/\partial T)_V$ and $P=-(\partial F/\partial V)_T$. Introducing the heat capacity at constant volume
	\be\label{heatcapacity}
	C_{V}=T\left(\frac{\partial S}{\partial T}\right)_{\!\!V}\ ,
	\ee
	the second derivatives appearing in \eqref{ruppeiner0} are
	\be\label{firstlawstep}
	-\left(\frac{\partial^{2} F}{\partial T^{2}}\right)_{\!V}=\left(\frac{\partial S}{\partial T}\right)_{\!V}=\frac{C_V}{T}\ ,\qquad
	\left(\frac{\partial^{2} F}{\partial V^{2}}\right)_{\!T}=-\left(\frac{\partial P}{\partial V}\right)_{\!T}\ .
	\ee
	The line element \eqref{ruppeiner0} then takes the form
	\be\label{ruppeiner2}
	d l^{2}=\frac{C_{V}}{T^{2}}\, d T^{2}-\frac{\left(\partial_{V} P\right)_{T}}{T}\, d V^{2}\ ,
	\ee
	which we use in what follows.

	The scalar (Ricci) curvature of a metric of the form \eqref{ruppeiner0} can be written, for general fluctuation variables, as\footnote{We have corrected a sign error in the corresponding expression, Eq.~(46) of \cite{wei2019b}.}
	\begin{align}\label{ricci1}
		R &=\frac{1}{2 C_{x}^{2}\left(\partial_{x} y\right)^{2}}\bigg(T\left(\partial_{x} y\right)\left[\left(\partial_{x} C_{x}\right)^{2}+\left(\partial_{T} C_{x}\right)\left(\partial_{x} y-T \partial_{T, x} y\right)\right]\\
		&+C_{x}\left[\left(\partial_{x} y\right)^{2}+T\left(\left(\partial_{x} C_{x}\right)\left(\partial_{x, x} y\right)-T\left(\partial_{T, x} y\right)^{2}\right)+2 T\left(\partial_{x} y\right)\left(-\left(\partial_{x, x} C_{x}\right)+T\left(\partial_{T, T, x} y\right)\right)\right]\bigg)\ ,\nonumber
	\end{align}
	the sign of which determines the nature of the underlying interactions. Following the convention of \cite{ruppeiner1995}, a {\em positive} $R$ indicates a predominantly {\em repulsive interaction} and corresponds locally to the geometry of a sphere; a negative $R$ means an {\it attractive interaction} and the geometry of a pseudosphere, and $R=0$ indicates a flat geometry with no net interaction, as for the ideal gas. In the Reissner--Nordstr\"om-AdS case the magnitude of the curvature scales with the correlation length, $R\propto\xi^{d}$, and diverges at a critical point \cite{wei2019b}. 
	
	Let us now apply this machinery to the STU black holes. We specialize \eqref{ricci1} to the fluctuation variables $x=V$ and $y=P$, with $T$ and $P$ defined in Section~\ref{stu}. Since we know explicitly the relations 
	\be 
	T=T(P,r_+)\,,\quad V=V(P,r_+)\ ,
	\ee 
	(see Eq.~\eqref{TDs}) then for any function $f=f(P,r_+)$ we can calculate 
	\be 
	\frac{\partial f}{\partial V}\Bigl|_{T}=\frac{\{f,T\}_{P,r_+}}{\{V,T\}_{P,r_+}}\equiv \frac{\frac{\partial f}{\partial P}\frac{\partial T}{\partial r_+}-\frac{\partial f}{\partial r_+}\frac{\partial T}{\partial P}}{
		\frac{\partial V}{\partial P}\frac{\partial T}{\partial r_+}-\frac{\partial V}{\partial r_+}\frac{\partial T}{\partial P}
	}\ ,
	\ee 
	where we have introduced the corresponding `Poisson bracket', and similarly 
	\be 
	\frac{\partial f}{\partial T}\Bigl|_{V}=\frac{\{f,V\}_{P,r_+}}{\{T, V\}_{P,r_+}}\ ,
	\ee  
	(see Appendix~\ref{A1} for more details). As before, we will first consider the Reissner--Nordstr\"om-AdS black hole, and then turn towards the single and doubly charged cases.
	
	\subsection{Reissner--Nordstr\"om-AdS}
	
	For the Reissner--Nordstr\"om-AdS case \eqref{TDsRN} the entropy and the thermodynamic volume are both functions of the horizon radius alone, so that $S=S(V)$ and the heat capacity at constant volume is zero:
	\be\label{cv}
	C_{V}=T\left(\frac{\partial S}{\partial T}\right)_{\!\!V}=0\,.
	\ee
	The metric \eqref{ruppeiner2} is then degenerate, and its curvature \eqref{ricci1} diverges. Following \cite{wei2019a, wei2019b}, we deal with this by renormalizing the curvature by a factor of $C_V$,
	\be\label{ricci2}
	R_N\equiv R\,C_V\ ,
	\ee
	which is finite as $C_V\to0$. Evaluating \eqref{ricci2} at constant $C_V$ then gives the renormalized curvature in terms of the equation of state alone,
	\be\label{ricci2b}
	R_N=\frac{\left(\partial_{V} P\right)^{2}-T^{2}\left(\partial_{V, T} P\right)^{2}+2 T^{2}\left(\partial_{V} P\right)\left(\partial_{V, T, T} P\right)}{2\left(\partial_{V} P\right)^{2}}\ .
	\ee
	The denominator is proportional to $(\partial_V P)^2$ so $R_N$ diverges precisely at the equation of state extrema, $\partial_V P=0$. 
	
	The sign of the curvature can, in this case, be determined analytically. Since $(\partial_T P)_V = 1/(2r_+)$ is independent of $T$, $\partial_{V,T,T} P$ vanishes, and (\ref{ricci2b}) reduces to \ba\label{RNequal}
	R_N &=& \frac{1}{2}\left( 1 - \left( \frac{T \, \partial_{V,T} P}{\partial_V P} \right)^{\!2} \right)=\frac{8(8Q^2-r_+^2)(2\pi P r_+^4+Q^2)}{(8\pi Pr_+^4+12Q^2-r_+^2)^2}\,\nonumber\\
	&=&\frac{64 \pi ^{4/3} \left(3 Q^2 T V+8 Q^4\right)-32 (6 \pi )^{2/3} Q^2 V^{2/3}-6 (6 \pi )^{2/3} T V^{5/3}+3 \sqrt[3]{6}\, V^{4/3}}{\left(6^{2/3} V^{2/3}-2 \pi ^{2/3}\left(16 Q^2+3 T V\right)\right)^{2}}\ .  
	\ea
	Obviously, $R_N \leq 1/2$ everywhere and it is unbounded below. When the pressure is positive, the sign of the Ruppeiner curvature is given by
	\begin{equation}
		R_N > 0 \quad \Leftrightarrow \quad r_+ < 2\sqrt{2}\,|Q| 
		\qquad\implies\qquad
		V < \frac{64\sqrt{2}\,\pi}{3} |Q|^3\ .
		\label{RNpositive}
	\end{equation}
	This region is always physical since the extremal radius is $r_{\rm e} = 2|Q|$ at $P= 0$ and decreases with pressure, so the band $r_{\rm e} < r_+ < 2\sqrt{2}\,|Q|$ has $T > 0$ at any positive pressure. This is the region where repulsive interactions were observed in \cite{wei2019b}. A second positive region does lie on the unphysical branch, at $r_+ > 2\sqrt{2}\,|Q|$ together with $P < -Q^2/(2\pi r_+^4)$. These are asymptotically de Sitter configurations and are removed once Maxwell's equal-area law is imposed.

\begin{figure}[h]
		\centering
		\centering    \includegraphics[width=0.85\textwidth]{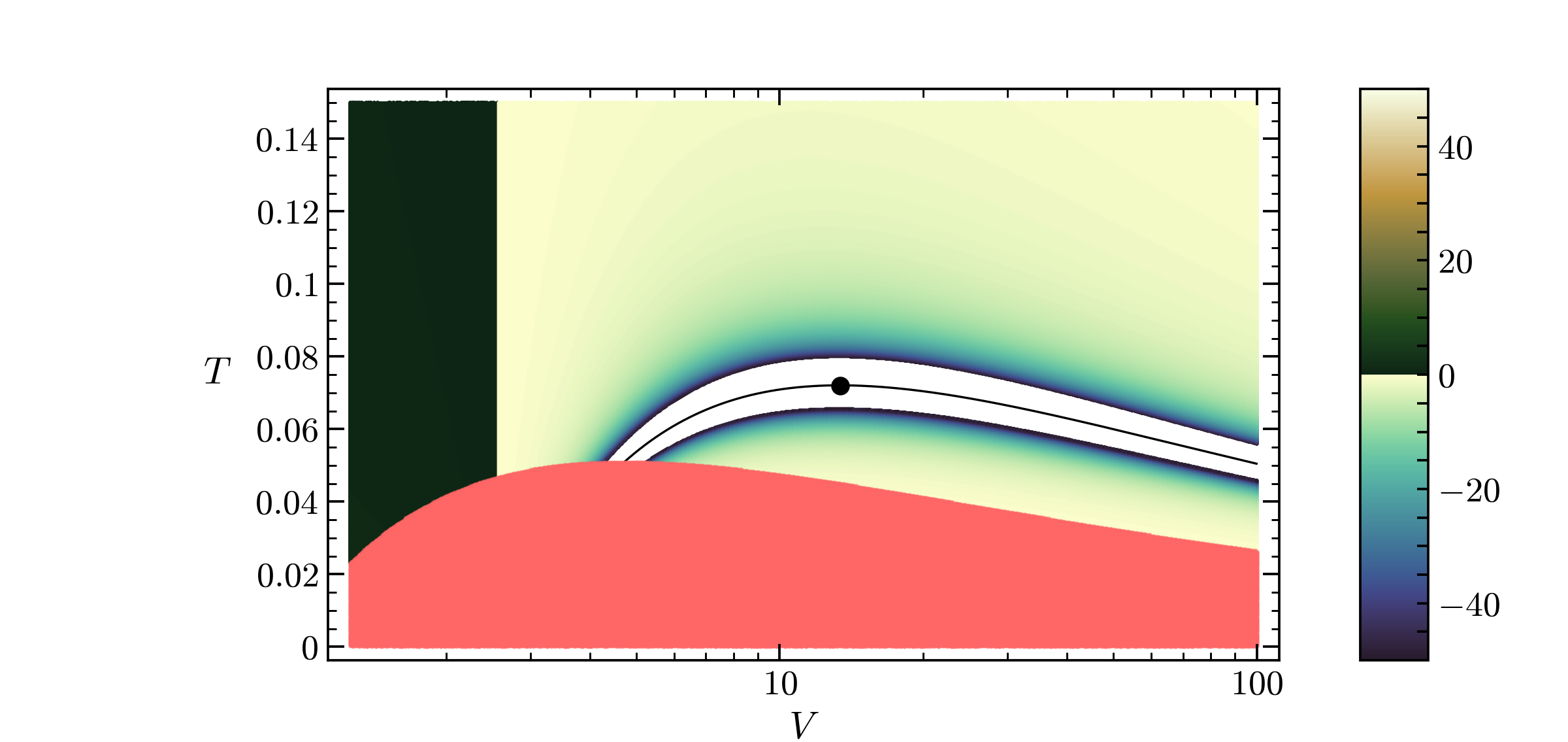}\hspace{1cm}
		\caption{{\bf Ruppeiner curvature $R_N$ for Reissner--Nordstr\"om-AdS.} The renormalized Ruppeiner curvature $R_N$ is displayed as a function of two fluctuation coordinates $(T,V)$ with $Q=0.3$. The black region corresponds to repulsive interactions, indicated by $R_N>0$. The red region is unphysical as it has $P<0$ and corresponds to asymptotically de Sitter black holes. The black curve represents the locus along which $(\partial_V P)_T=0$ where $R_N$ diverges. The black dot marks the critical point, where also $(\partial_V^2 P)_T=0$. The white region appears because the scale of $R_N$ is cut off at $R_N=\pm 50$.}
		\label{RN1}
	\end{figure}

	The renormalized curvature scalar \eqref{RNequal} is displayed in Fig.~\ref{RN1}. The positive $R_N$ region \eqref{RNpositive} corresponds to the black region on the left. The red region indicates negative pressures and is removed (cf. Fig.~\ref{fig1}). The black curve is the $(\partial_V P)_T=0$ line. It is not the coexistence line but instead represents the `boundary of local stability'. The coexistence line is fixed by Maxwell's equal area construction and lies strictly inside the region bounded by this curve, and the curvature is finite there. The $(\partial_V P)_T=0$ curve has two branches and points along them correspond to the two cusps in a $G-T$ diagram, which meet at the critical point (marked by the black dot) where $(\partial_V^2 P)_T$ vanishes as well. Above the critical temperature the curvature is finite everywhere and no phase transitions are present.

	The change in sign in $R_N$ implies a qualitative change in the microscopic interactions. Since it occurs predominantly at small radius, one interpretation is that the constituents there become dense enough for short-range repulsion to dominate. Such an interpretation rests on the expectation that the black hole possesses a genuine microstructure and that the sign of the Ruppeiner curvature faithfully reflects the interactions among its constituents.

	\subsection{Single charge and two unequal charges}
	
	In the single or two (unequal) charge cases, the heat capacity $C_V$ no longer vanishes. Instead, it is given by
	\be\label{cv2}
	C_V=T\left(\frac{\partial S}{\partial T}\right)_{\!V}
	=\frac{2\pi r_+\,T\,\partial_P V}
	{\partial_{r_+}T\,\partial_P V-\partial_P T\,\partial_{r_+}V}\ .
	\ee
	The numerator vanishes when $\partial_{P}V=0$, that is, when the thermodynamic volume is independent of the pressure at fixed $r_+$. For the Reissner--Nordstr\"om-AdS  case $V=\tfrac{4}{3}\pi r_+^3$ carries no pressure dependence, so $\partial_{P}V=0$ and $C_V=0$ identically, and the curvature must be renormalized as in the previous subsection.  For the single or unequal two charged cases, however, $V$ depends on $P$ (see Eq.~\eqref{singleTDs} and \eqref{TDs}, respectively) so  $\partial_{P}V\neq0$, and $C_V$ is nonzero. 
	
	Interestingly, while the corresponding expression for $C_V$ is in this case cumbersome and not illuminating to write down, one can check that $C_V>0$ whenever $P>0$ and $T>0$. This means that $R$ and $R_N=RC_V$ only differ in magnitude, and either quantity can be used to determine the repulsive/attractive character of interactions. In what follows, we choose to plot $R_N$. Note, however, that for its calculation, we can no longer use the simplified formula \eqref{ricci2b}, valid only for constant $C_V$, and have to employ  the full formula \eqref{ricci1}. 
	
	In the single charge case, there are no critical points.  Nevertheless, $R_N$ can still diverge. It happens when 
	\be 
	(\partial_VP)_T=0\ .
	\ee 
	Written out in terms of $r_+$ and $P$, this condition becomes a polynomial in the charge, which we give in Appendix~\ref{A2}, upon setting $Q_1=Q$ and $Q_2=0$. More importantly, $R_N$ is always negative, indicating attractive interactions in the entire parameter space for the single-charged case.

	\begin{figure}[h]
		\centering
		\centering    \includegraphics[width=0.85\textwidth]{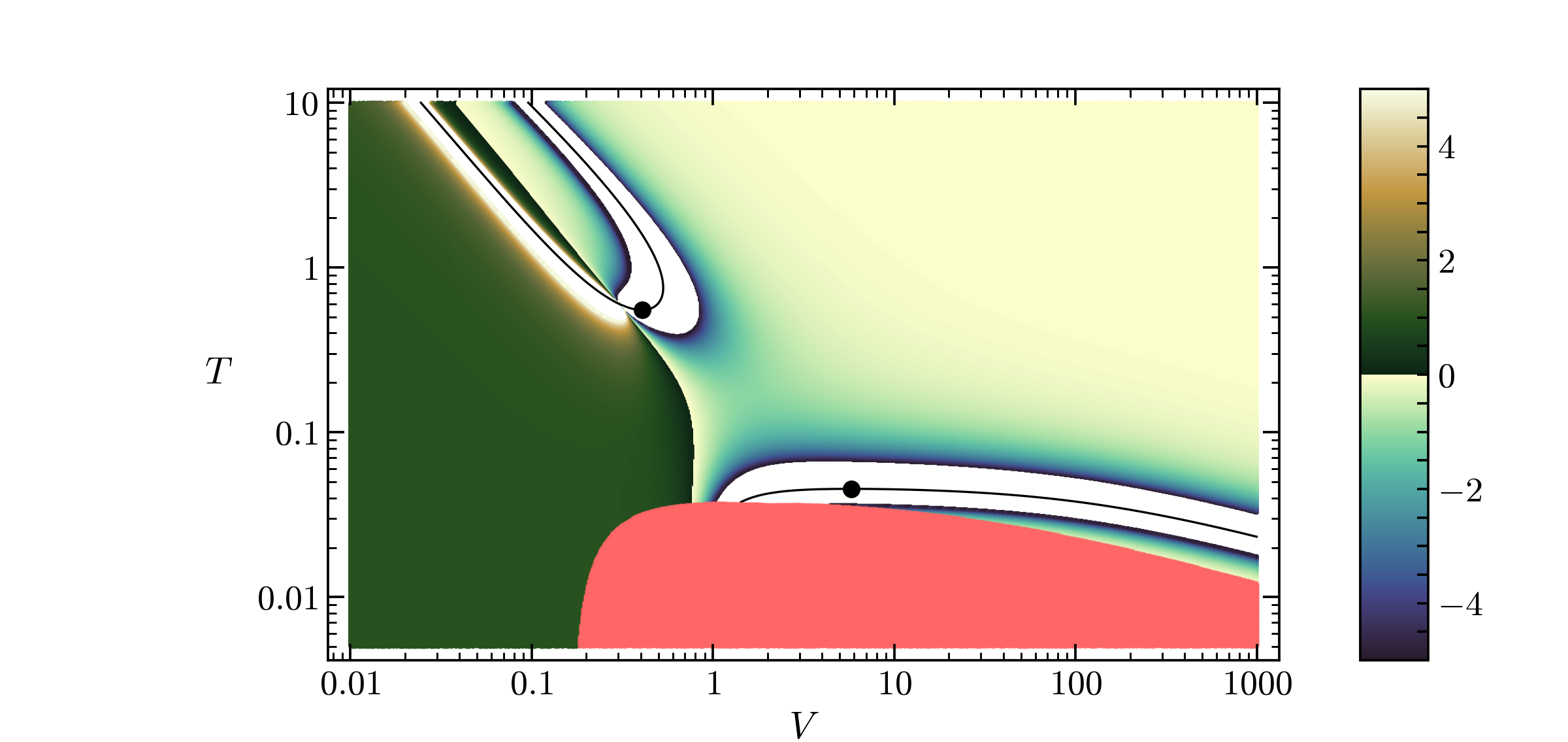}\hspace{1cm}
		\caption{{\bf Ruppeiner curvature $R_N$ in the unequal charge case.} The renormalized Ruppeiner curvature $R_N$ is displayed as a function of two fluctuation coordinates $(T,V)$ with $Q=1$ and $\epsilon=0.01$.  Red points indicate $P<0$ (de Sitter space). The black curve represents the locus along which $(\partial_V P)_T=0$, where $R_N$ diverges. The black dot marks the critical points, where also $(\partial_V^2 P)_T=0$. The reverse Van der Waals-like critical point is the one occurring at small volumes and large temperatures (upper left), whereas the standard critical point occurs at large volumes and small temperatures (right bottom). The $R_N=0$ curve is no longer straight and as $T$ increases, it  bends towards small volumes. Both critical points occur in the region where $R_N<0$, corresponding to attractive forces. The white region appears because the scale of $R_N$ is cut off at $R_N=\pm 5$.}
		\label{RNunequal}
	\end{figure}

	For the case of unequal charges, we plot $R_N$ in Fig.~\ref{RNunequal}. To highlight the difference with the Reissner--Nordstr\"om-AdS case, we have chosen a charge ratio for which two critical points exist. The low temperature, large volume critical point is the standard Van der Waals-like critical point CP$_1$ displayed at the bottom right of the figure, and the large temperature, small volume critical point is the reverse Van der Waals-like critical point CP$_2$ (upper left). Contrary to the single-charged case, once a second charge is added,  the region of positive Ruppeiner curvature appears at small volumes. However,  contrary to the Reissner--Nordstr\"om-AdS case, the $R_N=0$ curve is no longer `straight' (at constant volume), but slightly bends towards smaller volumes as the temperature increases. Consequently, both critical points remain in the attractive interaction region, characterized by $R_N<0$. This is apparent for tiny $\epsilon$ values, while as $\epsilon$ increases (and the second critical point moves to infinity), the $R_N=0$ curve moves closer to the second critical point, which then appears to almost lie on this curve.

	\section{Conclusions}\label{conclusions}

	We have studied the thermodynamics and thermodynamic geometry of the non-rotating, pair-wise-equal charge STU black hole of gauged $\mathcal{N}=8$, $SO(8)$ supergravity, working in the extended phase space in which the cosmological constant is treated as a pressure and the mass as an enthalpy. We obtained closed-form expressions for its thermodynamic quantities directly in terms of the physical charges $Q_1$ and $Q_2$. This reformulation makes the equal-charge case transparent, as the solution reduces to that of the Reissner--Nordstr\"om-AdS black hole, and the equation of state and its critical point follow analytically. Once the charges are unequal, the mass acquires an irreducible square root, the analysis becomes intrinsically non-analytic, and the critical points are accessible only numerically.
	
	The thermodynamic behavior of the distinct-charge solution depends crucially on the charge ratio. Whereas it is Reissner--Nordstr\"om-AdS-like for numerically similar charges, when the two charges are sufficiently different, a second critical point appears. In the Gibbs free energy, this is reflected by the existence of a second, reversed swallowtail of the reverse Van der Waals type, and the associated phase diagram contains a critical region where no phase transitions occur, which is enclosed between the sub-critical and super-critical regions where first-order phase transitions take place between small and large black hole phases, similar to what was observed in \cite{frassino2014, dolan2014,  hu2024} for higher-curvature gravity black holes. 
	
	Perhaps the most interesting finding is the `motion' of the position of the two critical points as the charge ratio is varied. Namely, as the charge ratio increases from zero, the second critical point recedes to infinity, is absent for qualitatively comparable charges, and eventually reappears as the ratio increases even further. This is accompanied by a monotonic motion of the first critical point, which steadily moves towards smaller temperatures and pressures. 
	
	We have also constructed the Ruppeiner geometry for the STU black hole, using temperature and thermodynamic volume as the fluctuation variables, and examined its renormalized curvature as a probe of the underlying microstructure. In the equal-charge limit, we recovered the Reissner--Nordstr\"om-AdS results of \cite{wei2019a, wei2019b}, properly indicating the unphysical region of positive Ruppeiner  curvature that corresponds to asymptotically de Sitter states and must be excluded. Consequently, only the high temperature and small volume black holes can have positive Ruppeiner curvature and can feature repulsive interactions. One interpretation of the crossover region, which is above the extremal radius at zero pressure, indicates a scale at which the constituents become dense enough for  short range repulsion to dominate over the attraction governing the larger black holes.
	
	In the case of single charge solutions, there are no critical points, and the Ruppeiner curvature is always negative, indicating attractive interactions in the whole parameter space, similar to a Van der Waals fluid. For unequal charges, the sign of the Ruppeiner curvature cannot be located analytically, and we have relied on numerical methods.  The zero-crossing of $R_N$ is no longer at fixed volume, and as temperature increases, it bends towards lower volumes. Consequently, both critical points (if they exist) occur for $R_N<0$, indicating attractive interactions in those regions. In general, the introduction of a second charge yields repulsive interactions for low temperature black holes. 
	
	There are several natural extensions of this work to consider. The most immediate is the more general case of three or four distinct charges, where the analysis must be carried out numerically. The larger parameter space may support additional critical points or triple points. The inclusion of rotation, restoring the full Kerr--Newman-like solution (e.g. \cite{ma2026, wu2026}) is another natural step. 
	
	More interesting, in our opinion, is the holographic interpretation of the phase structure found here. These solutions descend from M-theory compactified on $AdS_4\times\mathcal{S}^7$, whose boundary dual is the superconformal field theory on a stack of M2-branes, and the four charges correspond to the Cartan generators of its $SO(8)$ $R$-symmetry. The two-charge phase diagram, with its second critical point and post-critical phases, should map to the thermal phase structure of the theory at finite $R$-charge chemical potential. It would be worthwhile to explore the dual of this phase space in that theory. Similarly, despite a lack of closed-form for the critical points, a numerical determination of the critical exponents should be possible, which would establish whether the second reversed critical point lies in the same mean-field universality class as the Van der Waals point or a distinct one. The presence of two critical points also makes the STU solution a natural testing ground for the recently developed thermodynamic topology of black hole critical phenomena 
	\cite{wei2022}.

	\section*{Acknowledgments} 
	
	We are grateful to Natacha Altamirano for preliminary calculations during the early stages of this project. D.K. acknowledges support from the Charles University Research Center Grant No. UNCE24/SCI/016. This work was supported in part by the Natural Sciences and Engineering Research Council of Canada, and in part by the Institut Philippe Meyer.

	\appendix
	
	\section{Technical results}

	\subsection{Calculating the Ruppeiner curvature}\label{A1}

	For multiple derivatives, we use a generalization of the Jacobian method, with equation (2.19) in \cite{mansoori2015}. In short, if we have variables $x, y$, functions of those variables $A(x,y) , B(x,y)$ (which will play the role of new coordinates) and any other new function $F(x,y)$, we can find that
	\be 
	\left(\frac{\partial F}{\partial A}\right)_B = \frac{\{F, B\}}{\{A,B\}}\ ,
	\ee 
	where we use Poisson brackets 
	\be 
	\{X, Y\} = \frac{\partial X}{\partial x}\frac{\partial Y}{\partial y}- \frac{\partial Y}{\partial x}\frac{\partial X}{\partial y}\ .
	\ee 
	For $F = x$ or $y$, we find the same result as with the Jacobian approach.

	Nesting these partial derivatives allows us to calculate the expressions in \eqref{ricci1}, without having to invert any function. We choose $x = P$, $y = r_+$, and from here we can use numerics to get an explicit formula for $R_N$, by doing, for example: 
	\be 
	\partial_{V,T} P = \frac{\{\frac{\{P, T\}}{\{V, T\}}, V\}}{\{T,V\}}\,.
	\ee 
	The expressions for $R_N$ quickly become very large and impractical to write down.

	\subsection{Divergence of the Ruppeiner curvature}
	\label{A2}
	
	The renormalized Ruppeiner curvature $R_N$ is given by \eqref{ricci2} and \eqref{ricci1}.  It diverges when
	\be
	\partial_V P=0\ ,
	\ee
	where the derivative is to be taken at fixed temperature. Therefore, we have
	\be
	\partial_V P=\left(\frac{\partial P}{\partial V}\right)_{\!T}=\frac{-\,\partial_{r_+} T}{\partial_{r_+} V\,\partial_P T-\partial_P V\,\partial_{r_+} T}\ ,
	\ee
	which, for a general two-charge case,  gives the following condition for the divergence of $R_N$:
	\begin{align}
		&768 Q_1^4 \Bigl(-4 \pi ^2 P^2 r_+^8+9 Q_2^2 \left(4 \pi  P r_+^4+r_+^2\right)+27 Q_2^4\Bigr)\nonumber\\
		&+96 Q_1^2 \Bigl(Q_2^2 r_+^4 \left(640 \pi ^2 P^2 r_+^4+288 \pi  P r_+^2+27\right)+72 Q_2^4 \left(4 \pi  P r_+^4+r_+^2\right)\nonumber\\
		&+2 \pi  P r_+^8 \left(8 \pi  P r_+^2+3\right)^2\Bigr)+r_+^8 \Bigl(4096 \pi ^3 P^3 r_+^4 \left(3 Q_2^2+r_+^2\right)-384 \pi ^2 P^2 (-24 Q_2^2 r_+^2+8 Q_2^4-3 r_+^4)\nonumber\\
		&+4096 \pi ^4 P^4 r_+^8+1728 \pi  P Q_2^2-27\Bigr)=0\,.
	\end{align}

	\subsection{Charge ratio for the second critical point}\label{A3}
	
	As discussed in the main text, for the two-charge STU black hole there is a range of charge ratios, for which only the usual Van der Waals critical point (denoted by $\mathrm{CP}_1$ in the main text) exists. This happens for the following charge ratios: 
	\be\label{tworoots} 
	\epsilon=\frac{Q_2}{Q_1}\in(\epsilon_c, 1/\epsilon_c)\,,\quad \epsilon_c=7-4\sqrt{3}=\frac{1}{7+4\sqrt{3}}\simeq\frac{1}{13.9282032}\simeq   0.0717968\,.
	\ee 
	Just below $\epsilon_c$ and above $1/\epsilon_c$, the second (reverse Van der Waals) critical point (denoted by $\mathrm{CP}_2$ in the main text) emerges and `descends' from infinite pressures and temperatures. It is the aim of this appendix to show how these analytical bounds on charge ratios emerge.

	To start with, we recall the two-charge temperature formula, \eqref{TDs}:
	\be\label{T} 
	T=\frac{r_+^4+4r_+^4g^2(Q_1^2+Q_2^2)+4g^2r_+^6+3g^4r_+^8-16Q_1^2Q_2^2}{4\pi r_+^3\chi_+}\,,
	\ee 
	where $\chi_+$ is given by \eqref{chiplus}.
	
	Since the thermodynamic volume is monotonic in $r_+$ at fixed $P=3g^2/(8\pi)$, the critical point conditions \eqref{crit}
	are equivalent to
	\begin{equation}
		\left(\frac{\partial T}{\partial r_+}\right)_{\!P} = 0\ ,
		\qquad
		\left(\frac{\partial^2 T}{\partial r_+^2}\right)_{\!P} = 0\ .
		\label{critdef}
	\end{equation}
	We assume $Q_1, Q_2 > 0$ with $Q_2 \geq Q_1$. At fixed pressure the first of the conditions \eqref{critdef} has either two positive roots or none, and the count can change only when two roots merge, which is where both conditions of \eqref{critdef} become true. Scanning the pressure upwards from zero, the count goes $2 \to 0$ at the ordinary critical point and, for sufficiently unequal charges, $0 \to 2$ again at a second and higher one. As the charge ratio is lowered towards its threshold value this second pressure grows without bound, so the threshold can be decided by the behaviour of \eqref{T} at large $P$.
	
	Let us now, in addition to $\epsilon=Q_2/Q_1$, introduce another dimensionless ratio (motivated by the second term in the numerator of \eqref{T}):
	\begin{equation}
		\sigma \equiv \frac{g^2r_+^4}{4Q_1^2}\,,
		\label{ksigma}
	\end{equation}
	and rewrite the temperature \eqref{T}, $T=T(r_+,P,Q_1,Q_2)$ in terms of $T=T(\sigma, P, Q_1, \epsilon)$, and perform an expansion $g\to\infty$
	$(P\to \infty$), while keeping $\sigma$ {\em fixed}. 
	This yields
	
	\begin{equation}
		T = \frac{g^{3/2}\sqrt{2|Q_1|}}{4\pi}\,f(\sigma)+O(\sqrt{g})\,,\quad 
		f(\sigma) \equiv \frac{3\sigma^2 + (1+\epsilon^2)\sigma - \epsilon^2}{\sigma^{3/4}\sqrt{(\sigma+1)(\sigma+\epsilon^2)}}\ .\label{f}
	\end{equation}
	
	Since $\sigma$ is increasing in
	$r_+$ at fixed $g$, stationary points in $r_+$ correspond to those in $\sigma$. So, the stationary points of the limiting temperature are those of $f$ and depend on the charges only through $\epsilon$. Differentiating gives
	\begin{equation}
		f'(\sigma)
		= \frac{\Phi(\sigma)}{4\,\sigma^{7/4}(\sigma+1)^{3/2}(\sigma+\epsilon^2)^{3/2}}\ ,
		\label{fprime}
	\end{equation}
	where
	\begin{equation}
		\Phi(\sigma)
		= 3\sigma^4 + 6(\epsilon^2+1)\sigma^3
		- \bigl(\epsilon^4 - 20\epsilon^2 + 1\bigr)\sigma^2 + 6\epsilon^2(\epsilon^2+1)\sigma + 3\epsilon^4\ .
		\label{Phi}
	\end{equation}
	The denominator of \eqref{fprime} is positive and finite for $\sigma > 0$, so now
	the stationary points of $f$ are  the positive roots of the quartic
	$\Phi$. We count the positive roots of $\Phi$ (as a function of $\epsilon$) by looking for when two roots merge. This occurs where the discriminant of $\Phi$ vanishes. It is
	\begin{equation}
		\Delta_\Phi
		= 2304\,\epsilon^4\,(\epsilon-1)^6(\epsilon+1)^6
		\bigl(\epsilon^2 - 14\epsilon + 1\bigr)\bigl(\epsilon^2 + 14\epsilon + 1\bigr)\ .
		\label{disc}
	\end{equation}
	The interesting zeros are given by
	\begin{equation}
		\epsilon^2 - 14 \epsilon + 1 = 0\ ,
	\end{equation}
	which has two roots, $\epsilon=7\pm 4 \sqrt{3}$, which precisely coincide with $1/\epsilon_c$ and $\epsilon_c$ given in \eqref{tworoots}.
	In particular, the number of critical points for $\epsilon>1$ changes when $\epsilon$ crosses the value $7+ 4 \sqrt{3}$, and we can evaluate it once in each interval to determine the number. At $\epsilon = 10$ the quartic has two real roots but both are negative. At $\epsilon = 20$ it has four, with two of them positive. Hence, $\Phi$ has a pair of positive roots when $\epsilon \geq 7+4\sqrt{3}$. A similar argument then holds for $\epsilon<\epsilon_c$.

	\bibliographystyle{JHEP}
	\bibliography{LBIB}
	
\end{document}